\documentclass[aps,pre,showpacs,showkeys,reprint,eqsecnum,floatfix]{revtex4-1}

\usepackage{amssymb}
\usepackage{amsmath}
\usepackage{graphicx}
\usepackage{amsbsy}
\usepackage{color}
\usepackage{bm}

\newcommand{\bq}{\begin{eqnarray}}
\newcommand{\eq}{\end{eqnarray}}
\newcommand{\bqn}{\begin{eqnarray*}}
\newcommand{\eqn}{\end{eqnarray*}}
\newcommand{\nn}{\mathbf{n}}
\newcommand{\rr}{\mathbf{r}}
\newcommand{\kk}{\mathbf{k}}

\newcommand{\gio}[1]{#1}

\begin{document}
\title{Monte Carlo simulation of the nonadditive restricted primitive
  model of ionic fluids: Phase diagram and clustering} 

\author{Riccardo Fantoni}
\email{rfantoni@ts.infn.it}
\affiliation{Dipartimento di Scienze dei Materiali e Nanosistemi,
  Universit\`a Ca' Foscari Venezia, Calle Larga S. Marta DD2137, I-30123
  Venezia, Italy}
\author{Giorgio Pastore}
\email{pastore@ts.infn.it}
\affiliation{Dipartimento di Fisica dell' Universit\`a and
  IOM-CNR, Strada Costiera 11, 34151 Trieste, Italy} 

\date{\today}

\begin{abstract}
We report an accurate Monte Carlo calculation of the phase diagram and
clustering properties of the restricted primitive model with
non-additive hard-sphere diameters. At high density the positively
non-additive fluid shows more clustering than the additive model and
the negatively non-additive fluid shows less clustering than the
additive model, at low density the reverse scenario appears. A
negative nonadditivity tends to favor the formation of neutrally
charged clusters starting from the dipole. A positive nonadditivity
favors the pairing of like ions at high density. The critical point
of the gas-liquid phase transition moves at higher temperatures and
higher densities for a negative nonadditivity and at lower
temperatures and lower densities for a positive nonadditivity. The law
of corresponding states does not seem to hold strictly. Our results
can be used to interpret recent experimental works on room temperature
ionic liquids.
\end{abstract}

\maketitle
\section{Introduction}
\label{sec:introduction}

Ionic soft matter \cite{Henderson2004} is a class of conventional
condensed soft matter
\cite{deGennes1992,Fantoni04b,*Fantoni06a,*Fantoni10a,*Fantoni11a,*Fantoni12a,*Fantoni12c}
with prevailing contribution from electrostatics, 
in many cases crucially affecting
its physical properties. Among the most popular representatives of
such a class of materials are natural and synthetic saline
environments, like aqueous and non-aqueous electrolyte solutions 
and molten salts as well as a variety of polyelectrolytes and colloidal
suspensions. Equally well known are many biologically important
proteins.

The simplest theoretical model for ionic colloidal suspensions is the
Restricted Primitive Model (RPM) \cite{Hansen}, a binary mixture of
uniformly charged Hard-Spheres (HS) of diameter $\sigma$: two
species of opposite charge $\pm q$ and equal concentrations to ensure 
charge neutrality, 
moving in a medium of fixed dielectric constant $\epsilon$. The
phase diagram properties of this model have been widely studied both
through analytical theories 
\cite{Stell1976,Croxton1979,Gillan1983,Cartailler1992,Given1992,Given1992b,Fisher1993,Fisher1994,Stell1995,Zhou1995,Yeh1996,Given1997,Jiang2002}
and within computer experiments starting from the seminal works of Larsen
\cite{Larsen1979} and Vel'Yaminov
\cite{Veliaminov1976,Chasovshikh1976}, followed by the pioneering 
Gibbs Ensemble Monte Carlo (GEMC) calculation of Panagiotopoulos
\cite{Panagiotopoulos92} and by other numerical
simulations 
\cite{Graham1990,Caillol1994,Orkoulas1994,Caillol1995,Orkoulas1999,Yan1999,Luijten2002,Caillol2002}.
The more general primitive model (PM) with asymmetry in ion
charge \cite{Camp1999}, 
in ion size \cite{Enrique2000,Panagiotopoulos2002,Yan2001} and
in both \cite{Panagiotopoulos2002,Yan2002} has also been studied.

From these studies emerged how, in the vapor phase, an important role
is played by association and clustering. In an old paper
\cite{Pastore1985} one of us studied a modified RPM fluid where one
allows for size nonadditivity particle diameters. Controlling the
nonadditivity, it was suggested through the use of integral equation
theories, that such a fluid might have a complex behavior due
to the possible competition between clustering tendence due to the Coulomb
interaction and demixing tendence due to entropic advantage driven by the
nonadditivity. Thus, the nonadditivity of the hard-sphere diameters does
not destroy the simplifying symmetry of the model but  it
enriches the properties of the pure RPM model 
making it a paradigm for the self-assembly of isotropic
particles and a challenge to present day theories of fluids.
In real systems, the degree of nonadditivity might be
directly related to the anion-cation contact-pairing affinity
\cite{Kalcher2010} which in turn may be mediated by the solvent.

It is the purpose of this paper to reconsider such a model fluid from
the point of view of accurate numerical experiments. In particular, we
want to study the clustering properties of the fluid outside of the
gas-liquid coexistence region. To this aim we first determined the
gas-liquid coexistence curve through the Gibbs ensemble method after having
studied semi-quantitatively how the coexistence region changes with
the nonadditivity through a density distribution analysis in the
canonical ensemble. This way we
could be sure that our cluster analysis falls outside the coexistence
region in all the cases studied. Clustering turns out to be greatly
affected by the nonadditivity parameter. The
most striking effect being the prevalence of neutrally charged clusters
made up of an even number of particles in the negatively non-additive
fluid. When the nonadditivity allows complete overlap of the two
species of particles the formation of a fluid of neutral hard-spheres
of half the density is expected and our simulation results clearly show this
behavior. On the other hand, for a positive nonadditivity, it is known
that the neutral HS mixture tends to demix the two species and the
demixing critical density lowers as the nonadditivity increases
\cite{Rovere1994,*Lomba1996,*Jagannathan2003,*Gozdz2003,*Buhot2005,*Santos2010}.
We expect this property \gio{ of the neutral system 
to have some interesting effect on the clustering
properties of the charged fluid since demixing cannot occur in a
binary charged system}: The 
frustrated tendency to segregation of like particles and the reduced
space available to the ions favors pairing of like ions and
percolating clusters at high densities. To the
best of our knowledge this is the first time that such a model fluid
is studied with numerical simulations. Preliminary results from our
analysis have been presented in a letter \cite{Fantoni2013}, here we
extend that analysis and present for the first time the gas-liquid
binodal of the fluid as a function of the nonadditivity parameter.  

We think that the model fluid considered in this paper may be realized
experimentally through a colloid-star polymer mixture where both
species are charged \cite{Poon2001,*Poon2002} or by room temperature
ionic liquids
\cite{Weingartner1995,Kleemeier1999,Saracsan2006,Schroer2009} as 
discussed in Section \ref{sec:gemc}. In particular in the latter
systems liquid-liquid binodals shifted above and 
below the one of the pure RPM are observed depending on the kind of
solvent used. If on the one hand this can be ascribed to the different
dielectric constant of the solvent \cite{Kleemeier1999}, on the other
hand it is clear that, depending on the kind of solvent, the
anion-cation contact-pairing affinity may vary \cite{Kalcher2010} and
thus the different experimental ionic liquids should be
more correctly described by comparison not just with the pure RPM but
with the more realistic primitive model with the addition of either a
positive or negative size nonadditivity.  

The paper is organized as follows: in Section \ref{sec:model} the
model for the fluid we want to study is described, in
Section \ref{sec:results} the results from the numerical experiments are
reported. These are divided in a cluster analysis in Section
\ref{sec:clusters}, in an analysis of the radial distribution 
function and structure factor in Section \ref{sec:rdf}, and in an
analysis of the gas-liquid coexistence in Section
\ref{sec:binodal}. Theoretical remarks on the clustering properties
are presented in Section \ref{sec:theory} and Section
\ref{sec:conclusions} is for final remarks.  

\section{The model}
\label{sec:model}

The model fluid we want to study is the restricted primitive model
(RPM) of non-additive hard-spheres (NAHS). The RPM consists of $N/2$
uniformly charged hard-spheres of species 1 of diameter $\sigma$
carrying a total charge $+q$ each and $N/2$ uniformly charged
hard-spheres of species 2 of the same
diameter carrying a total charge $-q$ each. The spheres are moving in a
dielectric continuum of dielectric constant $\epsilon$ \gio{independent on the 
thermodynamic state}. The
interaction between an ion of species $i$ and one of species $j$ a
distance $r$ apart is given by 
\bq
\beta\phi_{ij}(r)=\left\{\begin{array}{ll}
+\infty & r\le \sigma_{ij}\\
\displaystyle
\frac{q_iq_j}{k_BT\epsilon r} & r>\sigma_{ij}
\end{array}\right.~,~~~i,j=1,2~,
\eq
where $\beta=1/k_BT$ with $T$ the absolute temperature and $k_B$ the
Boltzmann's constant, $q_i$ the charge of an ion of species $i$.
The ions form a mixture of non-additive hard-spheres, i.e.
\bq
\sigma_{ij}=\left\{\begin{array}{ll}
\sigma & i=j\\
\sigma (1+\Delta) & i\neq j
\end{array}\right.~,~~~i,j=1,2~,
\eq
with the nonadditivity parameter $\Delta>-1$ . A thermodynamic state
is completely specified by the reduced density
$\rho^*=\rho\sigma^3=N\sigma^3/V$, where $V$ is the volume containing
the fluid, 
the reduced temperature $T^*=k_BT\epsilon\sigma/q^2$
($q^2/(\epsilon\sigma)$ is taken as unit of energy), and the
nonadditivity parameter $\Delta$. We will call $x_1=\rho_-/\rho=1/2$
and $x_2=\rho_+/\rho=1/2$ the anions and cations molar concentrations
respectively. 

\section{Results}
\label{sec:results}

\begin{figure}[htbp]
\begin{center}
\includegraphics[width=8cm]{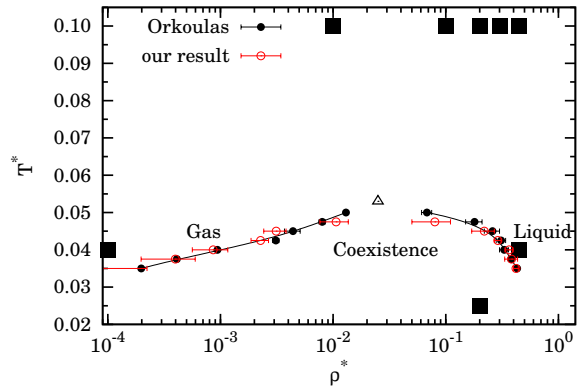}
\end{center}  
\caption{(Color online) Diagram showing the gas-liquid coexistence
  curve of the RPM 
fluid from the Gibbs ensemble MC data of Ref. \cite{Orkoulas1994}
(filled circles) and ours (open circles), the triangle being the
critical point, and the points (filled squares) of the phase diagram
where we run our $NVT$ MC simulations.}
\label{fig:coex}
\end{figure}

In Fig. \ref{fig:coex} we show the phase diagram of the pure RPM
fluid, $\Delta=0$, as obtained from the Gibbs ensemble Monte Carlo
method by Orkoulas {\sl et al.} \cite{Orkoulas1994} and by us (see
Section \ref{sec:gemc}). The thermodynamic points where we probe
the fluid with our $NVT$ Monte Carlo simulations are also shown as filled squares. 

In our canonical $NVT$ Monte Carlo (MC) we study the
fluid in a simulation box of volume $V=L^3$ with periodic boundary
conditions. The long range of the $1/r$ interaction is accounted for
using an Ewald sum for the interacting energy in the periodic
system \cite{Allen}. The interaction energy per unit box for
$\epsilon=1$ is calculated as 
\bq \nonumber 
{\cal U}&=&\sum_{\mu<\nu}\sum_\nn q_{i_\mu}q_{j_\nu}
\frac{\mbox{erfc}(\kappa|\rr_{\mu\nu}+L\nn|)}{|\rr_{\mu\nu}+L\nn|}\\ \nonumber
&&+\frac{2\pi}{L^3}\sum_{\kk\neq \mathbf{0}}\frac{e^{-(k/2\kappa)^2}}{k^2}|\rho_\kk|^2
-\frac{\kappa}{\sqrt{\pi}}\sum_\mu q_{i_\mu}^2\\ \label{energy}
&&-\frac{\pi}{2\kappa^2L^3}\left(\sum_\mu q_{i_\mu}\right)^2~,
\eq
where a Roman index with a Greek sub-index denotes the species of the
particle labeled by the Greek sub-index, $\rr_{\mu\nu}=\rr_\nu-\rr_\mu$
with $\rr_\mu$ the position of particle $\mu$, $\rho_\kk=\sum_\mu
q_{i_\mu}e^{-i\kk\cdot\rr_\mu}$, erfc denotes the 
complementary error function, $\nn=(n_x,n_y,n_z)$ with $n_x,n_y,n_k=0,\pm
1,\pm 2,\ldots$, and $\kk=(2\pi/L)(n_x,n_y,n_z)$ are
reciprocal lattice vectors. The parameter $\kappa$ which governs the
rate of convergence of the real space and reciprocal space
contribution to the energy is taken to be $\kappa\sim 5/L$. With this
value of $\kappa$, the real space contribution can be restricted to
the first term $\nn=\mathbf{0}$ only. The reciprocal space term includes
all $\kk$ vectors such that $n_x^2+n_y^2+n_z^2<27$. The last term in
Eq. (\ref{energy}) is zero for the RPM but it is important in the
Gibbs ensemble simulation where a particle exchange between the two
boxes can produce systems where there is an unequal number of positive
and negative charges. Our choice for the interaction energy takes into
account the fact that each charge has a uniform background of
neutralizing opposite charge density.

In our $NVT$ MC simulations we used $N=100$ (except for the test of
the size dependence of the clustering analysis where we considered up
to 5000 particles), the acceptance ratio is kept, on average,  
close to $50\%$ after a preliminary adjustment of the maximum particle
displacement. We start from a simple cubic configuration of two
crystals one made of species 1 and one made of species 2
juxtaposed in order to avoid overlaps at high densities. We need
around $10^5$ MC steps (MCS) in order to equilibrate the samples and
$10^6$ MCS$/$particle for the statistics.   

\subsection{Cluster analysis}
\label{sec:clusters}

During the simulation we perform a cluster analysis. After each 100
MCS we determine the number $N_n$ of clusters  made of
$n$ particles, so that $\sum_n nN_n=N$. We assume
\cite{Fantoni2011,*Fantoni2012} that a group of ions
forms a cluster if the distance $r$, calculated using
periodic boundary conditions, between a particle of species $i$
of the group and at least one other particle of species $j$ is
less then some fixed value, {\sl i.e.} 
$r<\sigma_{ij}+\delta^c\sigma$ where $\delta^c$ is a
parameter \cite{note1}. In all our simulations we choose $\delta^c=0.1$
(in Ref. \cite{Caillol1995} a detailed study of the sensitivity of the
clustering properties on this parameter is carried out for the pure RPM
fluid). Then we take the average of these numbers $\langle N_n\rangle$. 
Note that $Q_n=n\langle N_n\rangle/N$ gives the
probability that a particle belongs to a cluster of size $n$. 
To establish a criterion for percolation we first find the clusters 
without employing periodic boundary conditions to calculate the
distances, then we check whether, amongst the particles of any of
these clusters, there are two which satisfy the cluster condition
calculating the distances using periodic boundary conditions. Whenever 
we find one such cluster the cluster is percolating.    

\begin{figure*}[htbp]
\begin{center}
\includegraphics[width=8cm]{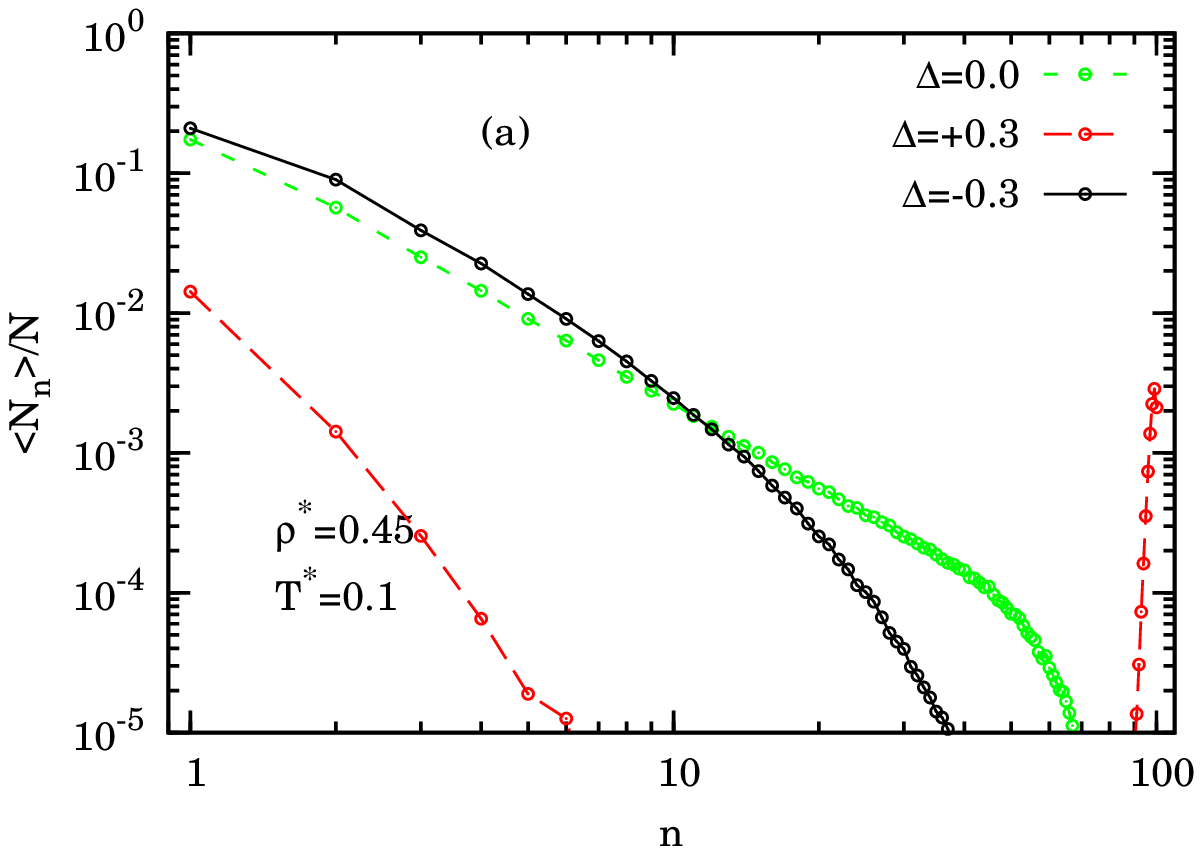}
\includegraphics[width=8cm]{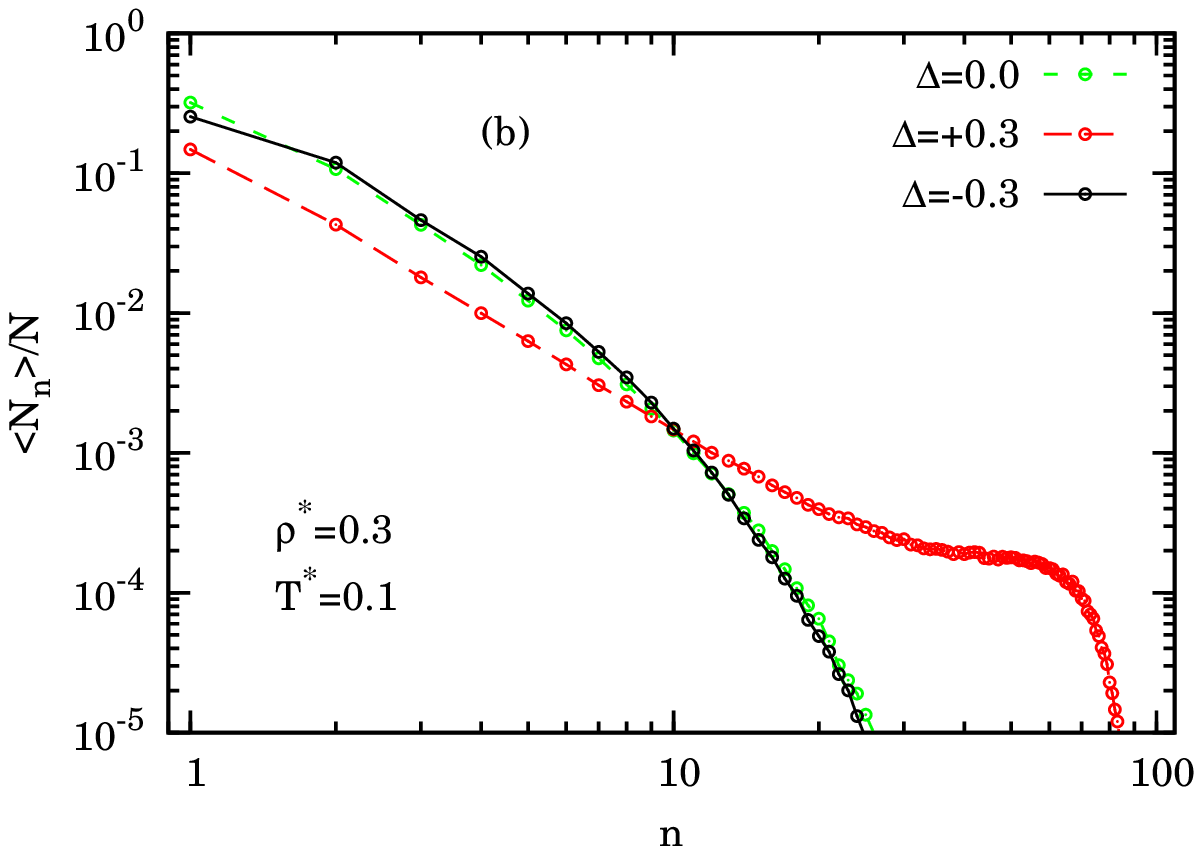}\\
\includegraphics[width=8cm]{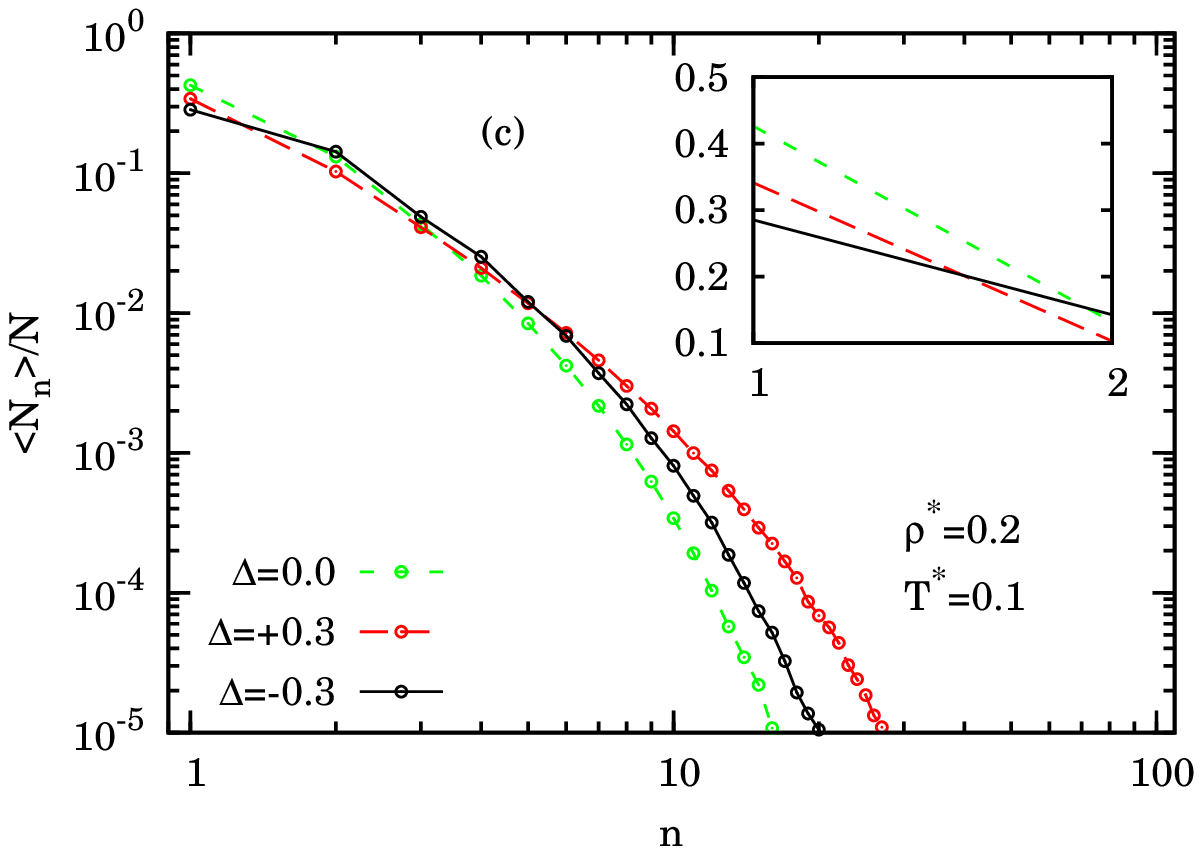}
\includegraphics[width=8cm]{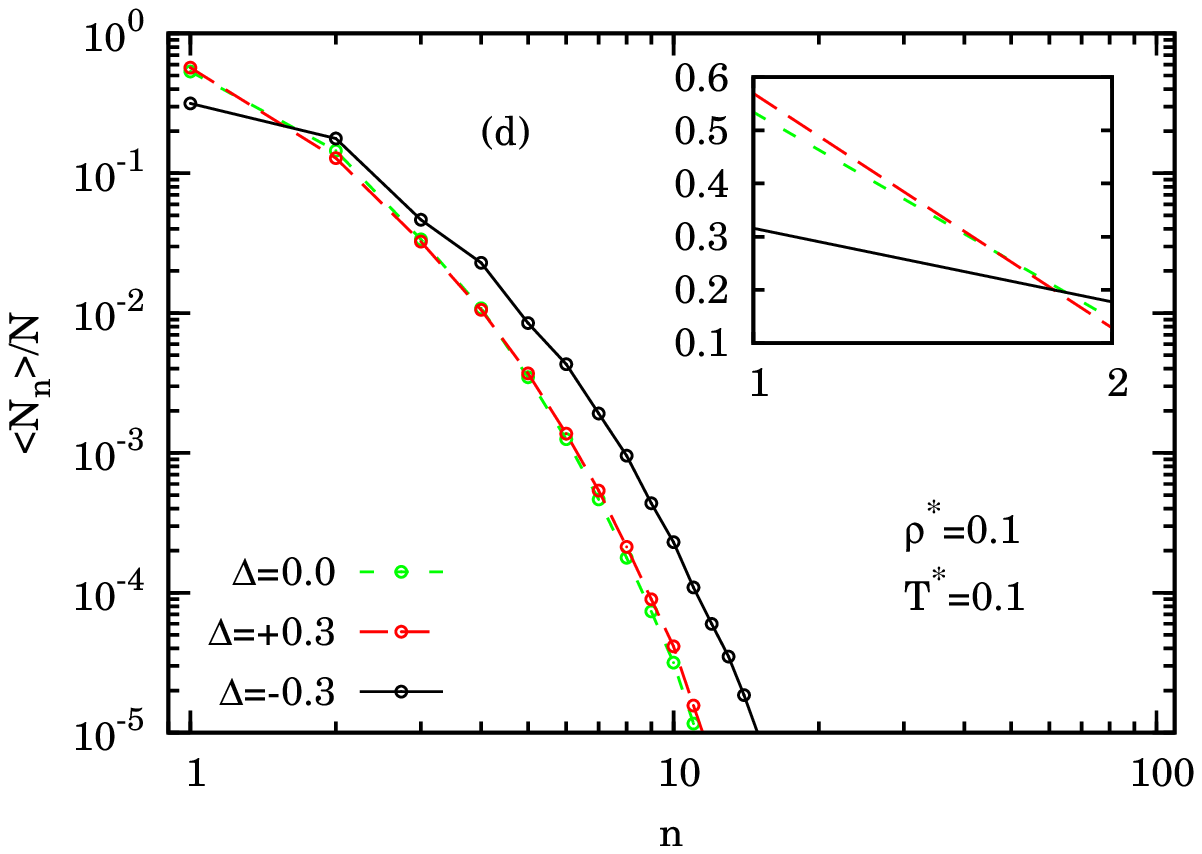}\\
\includegraphics[width=8cm]{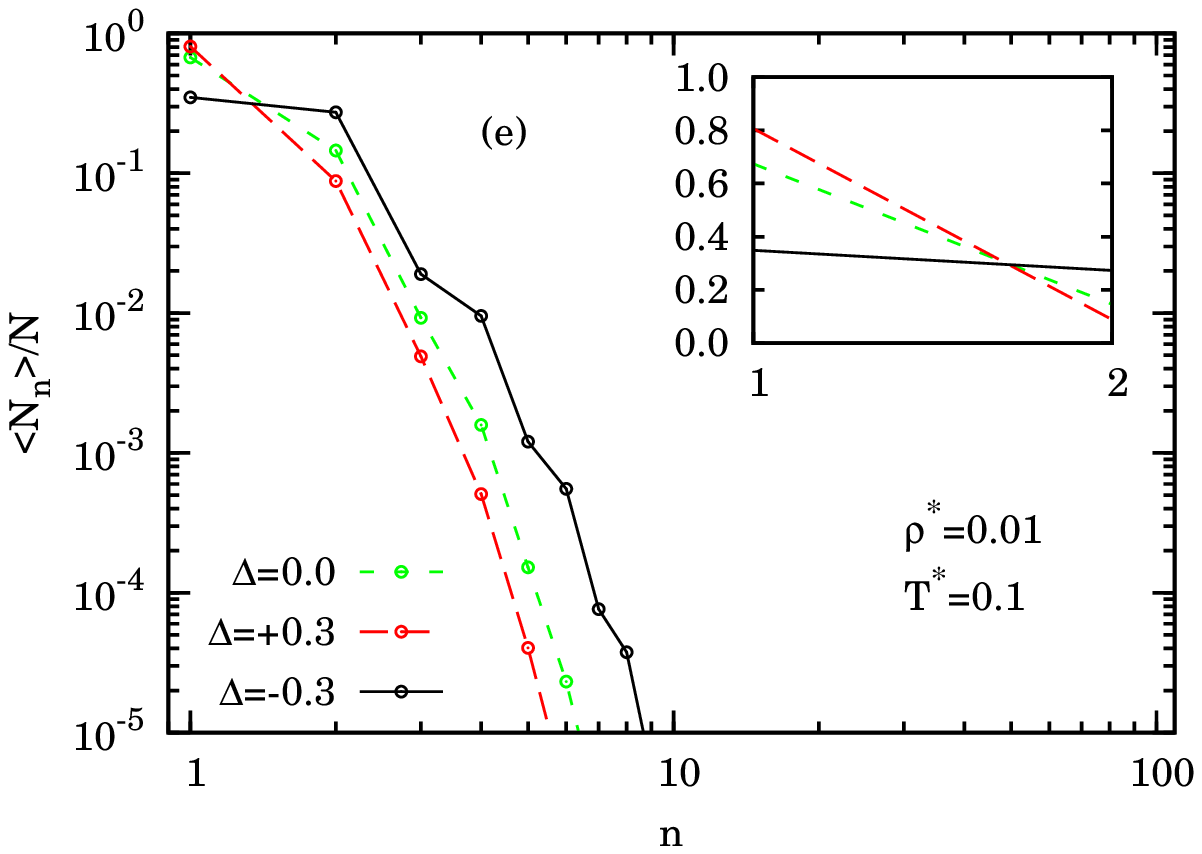}
\includegraphics[width=8cm]{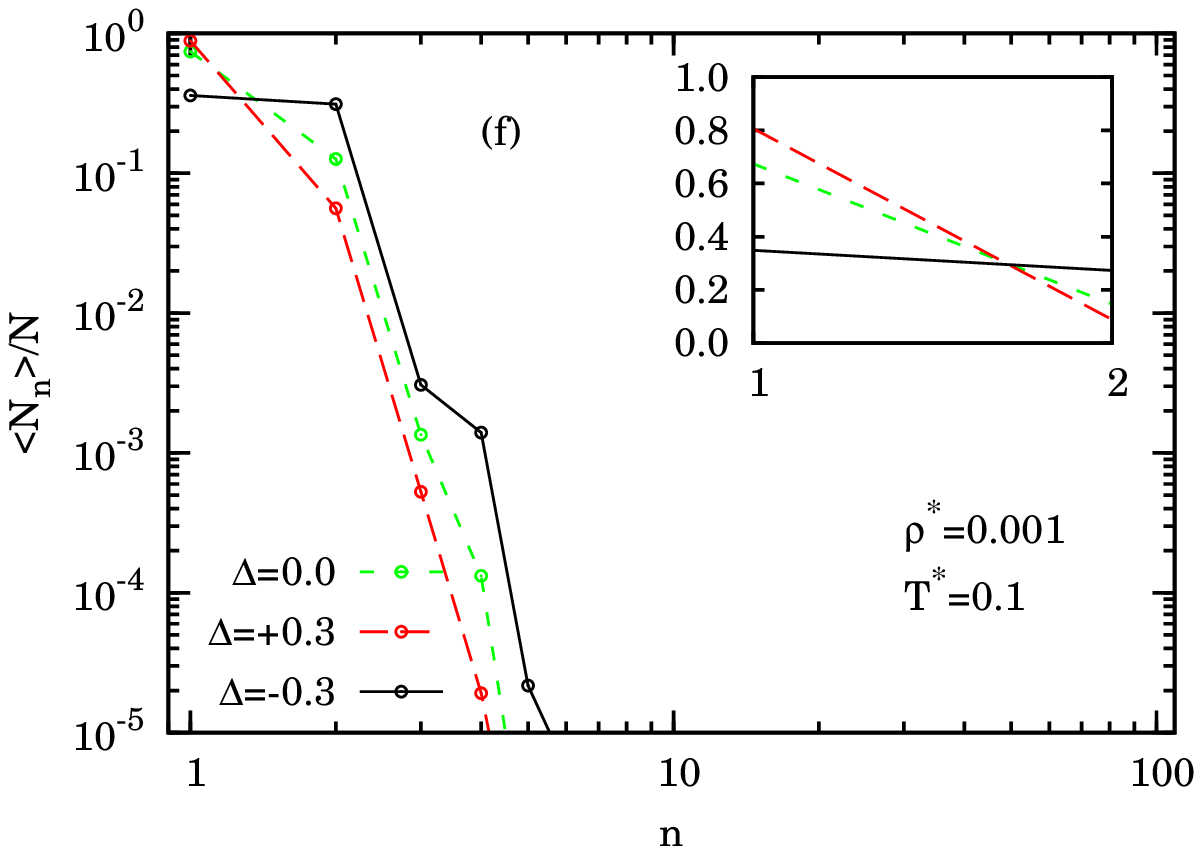}
\end{center}  
\caption{(Color online) Clustering properties of the fluid at
  $T^*=0.1$ at various 
values of nonadditivity. $N_n$ are the number of 
clusters made of $n$ particles. In the MC simulations we used $N=100$
particles and a number of MCS$=1\times 10^7$. The panels are ordered
(left to right, top to bottom) in order of decreasing density
$\rho^*=0.45, 0.3, 0.2, 0.1, 0.01$, and 
$0.001$ for panels (a), (b), (c), (d), (e), and (f) respectively. The
insets allows to read-off the degree of dissociation.}   
\label{fig:ncl-t0.1}
\end{figure*}

In Fig. \ref{fig:ncl-t0.1} we show results of such analysis
for the fluid at a temperature $T^*=0.1$ well above the critical
temperature, $T^*_c\approx 0.05$, of the 
pure RPM \cite{Orkoulas1999,Caillol2002,Luijten2002}. In the insets we
show a magnification of the region around $n=1$ from which the degree
of dissociation \cite{Zhou1995,Jiang2002} $\alpha=\langle N_1\rangle/N$
can be read-off. In the figure we plot the
cluster concentrations $\langle N_n\rangle/N$ as a function of the
number of particles $n$ in the cluster. We plot $n$ from $n=1$
(isolated ions) up to $n=N$ (in this case all the
particles of the fluid form one big percolating cluster). At
$\rho^*=0.45$ both the pure RPM and the $\Delta=+0.3$ fluid form 
percolating clusters. Lowering the density we first reach 
a state, at $\rho^*=0.3$, where the negative nonadditivity gives the
same clustering of RPM and the positive nonadditivity gives bigger
clustering (still with percolating clusters), then a state, at
$\rho^*=0.1$, where 
the positive nonadditivity gives the same clustering of RPM and the
negative nonadditivity a bigger one, and finally a state
$\rho^*=0.01, 0.001$, at low densities where a negative 
nonadditivity increases the clustering over the RPM fluid and a
positive nonadditivity diminishes it. Generally, at high densities we
find percolating clusters in the fluids whereas these disappear at low
densities even at a positive nonadditivity. 
Summarizing, in agreement with Ref. \cite{Pastore1985},
we find, for the fixed values of $|\Delta|$, that: 
at high density and positive $\Delta$ we have more clustering than
in the additive model since there is a smaller effective volume for 
particle, at high density and negative $\Delta$ we 
have less clustering than in the additive model because there is more
effective volume for the particles, at low density and positive $\Delta$ we
have less clustering than in the additive model due to the competition
between tendency to demixing in the corresponding neutral mixture and
tendency to local electroneutrality of the Coulombic systems, 
at low densities and negative $\Delta$ we have more
clustering than in the additive model because neutral clusters are
favored as shown in the next Section. We conclude that
at high temperature and high density the negative nonadditivity gives
lower clustering than RPM, lowering the temperature at constant
density or lowering the density at constant temperature it gradually
tends to gives higher clustering than RPM. On the contrary, at low
density the positive nonadditivity gives lower clustering than RPM,
increasing the density it gradually tends to give larger clustering
than RPM.

We determine the size dependence of the curves shown in
Fig. \ref{fig:ncl-t0.1} and see that when we have no percolating
clusters, for example the data at $T^*=0.1, \rho^*=0.3, \Delta=0,
-0.3$, the curves were unaffected by a choice of a higher number of
particles, while when we have percolating clusters, for example the
data at $T^*=0.1, 
\rho^*=0.3, \Delta=0.3$, the curve $(n,\langle N_n\rangle/N)$
changes with $N$. In these latter cases we find that a common   
curve is given by $(x,\langle N_x\rangle/N)$ with
$x=n/N\in [0,1]$. Then, in order to satisfy the normalization
condition,     
$1=\sum_nn(\langle N_n\rangle/N)\approx\int dx\,xN^2(\langle
N_x\rangle/N)$, we must have for two different sizes $N^\prime$ and
$N^{\prime\prime}$ 
that $(\langle N_x\rangle/N^\prime)/(\langle
N_x\rangle/N^{\prime\prime})\approx (N^{\prime\prime}/N^\prime)^2$. We
have no general recipe to when 
the former behavior is to be expected over the latter. We can only say
that the first behavior is generally observed when we do not have
percolating clusters whereas the second is present when we have
percolating clusters. In Section \ref{sec:theory} we show that the size
independent curves that we find when there are no 
percolating clusters can be fitted by $\langle N_n\rangle/N=a^nn^{bn}/n!$
(see Eq. (\ref{xcn}) with $z_n^{intra}$ obtained from an ideal
cluster approximation) with $a$ and $b$ a positive fitting
parameter. In Table \ref{tab:ab} we show the fitting parameters $a$ and
$b$ corresponding to the simulated cases.

\begin{table}[htbp]
\caption{Fitting parameters $a,b$ in the least square fit $\langle
  N_n\rangle/N=a^nn^{bn}/n!$ for the simulation results of
  Fig. \ref{fig:ncl-t0.1} without percolating clusters (and with the
  exclusion of the non smooth data at 
  $\rho^*=0.001$). The reduced $\chi^2$ was around $0.5$ with
  greater error approaching $n=1$. Also
  shown is the number of particles $n_{max}$ in the biggest cluster
  formed in each simulation.}   
\label{tab:ab}
{\scriptsize
\begin{center}
\begin{tabular}{|c|c||c|c|c|}
\hline
$\rho^*$ & $\Delta$ & $a$ & $b$ & $n_{max}$ \\ 
\hline
0.45&-0.3&0.220(3)&1.074(4) & 64 \\
0.3 &0   &0.197(4)&1.084(6) & 45 \\
0.3 &-0.3&0.204(3)&1.069(5) & 43 \\
0.2 &0   &0.206(7)&1.00(1)  & 23 \\
0.2 &+0.3&0.200(4)&1.083(5) & 45 \\
0.2 &-0.3&0.204(7)&1.04(1)  & 31 \\
0.1 &0   &0.22(2) &0.86(3)  & 15 \\
0.1 &+0.3&0.16(1) &1.01(4)  & 19 \\
0.1 &-0.3&0.15(1) &1.11(2)  & 29 \\
0.01&0   &0.41(7) &0.1(1)   & 8  \\
0.01&+0.3&0.36(8) &0.0(2)   & 7  \\
0.01&-0.3&0.23(4) &0.72(7)  & 12 \\   
\hline
\end{tabular}
\end{center}
}
\end{table}

\begin{figure}[htbp]
\begin{center}
\includegraphics[width=8cm]{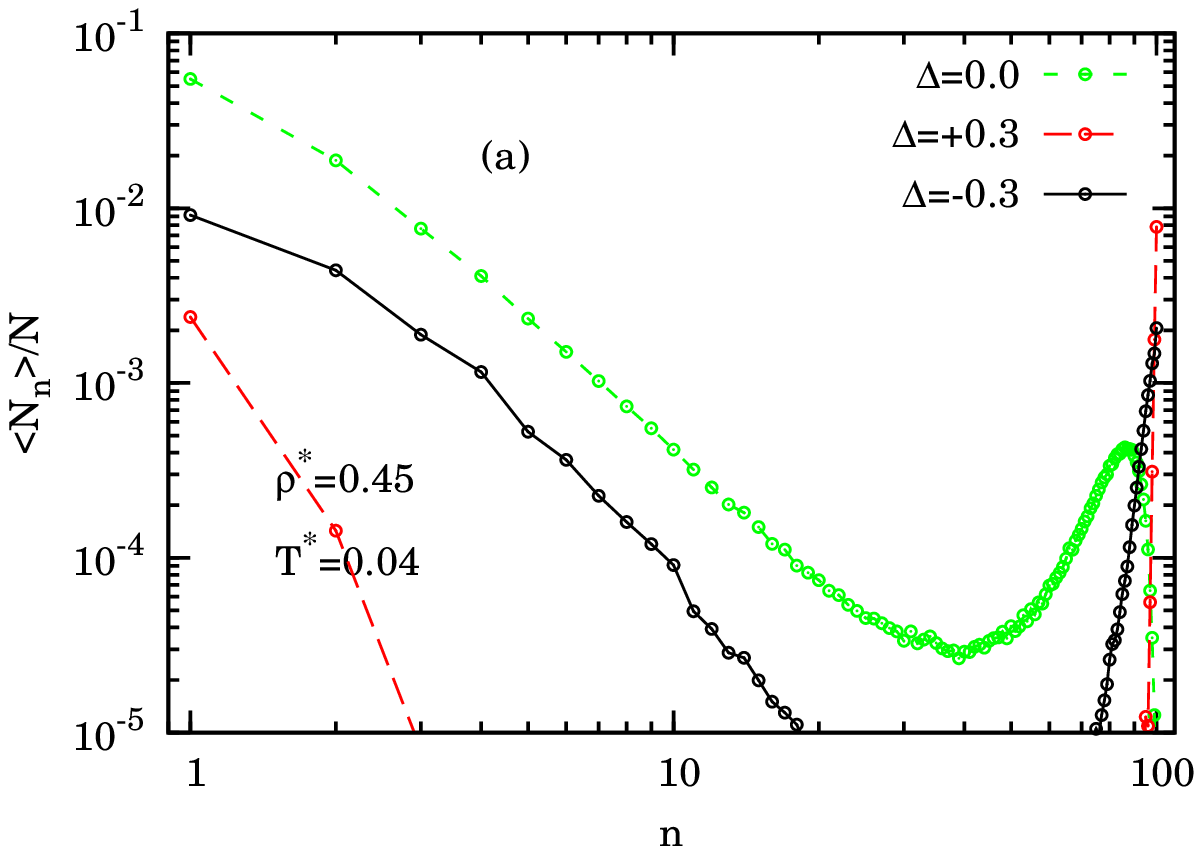}
\includegraphics[width=8cm]{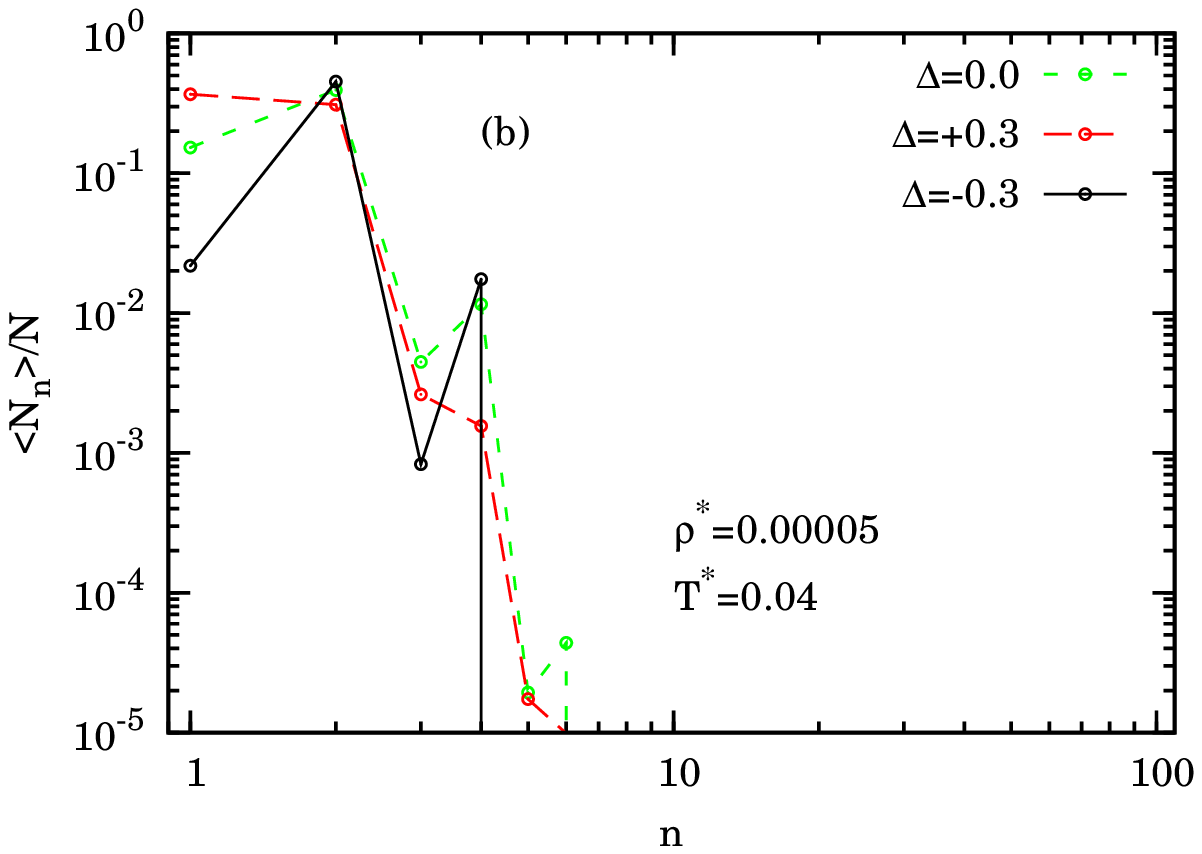}
\end{center}  
\caption{(Color online) Clustering properties of the fluid at
  $T^*=0.04$ and $\rho^*=0.45, 5\times 10^{-5}$, for panels (a) and
  (b) respectively, at various values of nonadditivity. $N_n$ are the
  number of clusters made of $n$ particles. In the MC simulations we
  used $N=100$ particles and a number of MCS$=1\times 10^7$.}
\label{fig:ncl-t0.04}
\end{figure}

In Fig. \ref{fig:ncl-t0.04} we show the clustering analysis at the
thermodynamic state below the critical temperature of RPM $T^*=0.04$
in the gas phase, $\rho^*=5\times 10^{-5}$, and in the liquid phase,
$\rho^*=0.45$. We see how in the gas phase only the first few
clusters are present in agreement with similar results found in
Ref. \cite{Caillol1995} and for a negative nonadditivity the dipoles
are clearly the preferred kind of clusters with the smallest
degree of dissociation amongst the three fluids considered. In the
liquid phase all three fluids have percolating clusters.

\begin{figure}[htbp]
\begin{center}
\includegraphics[width=8cm]{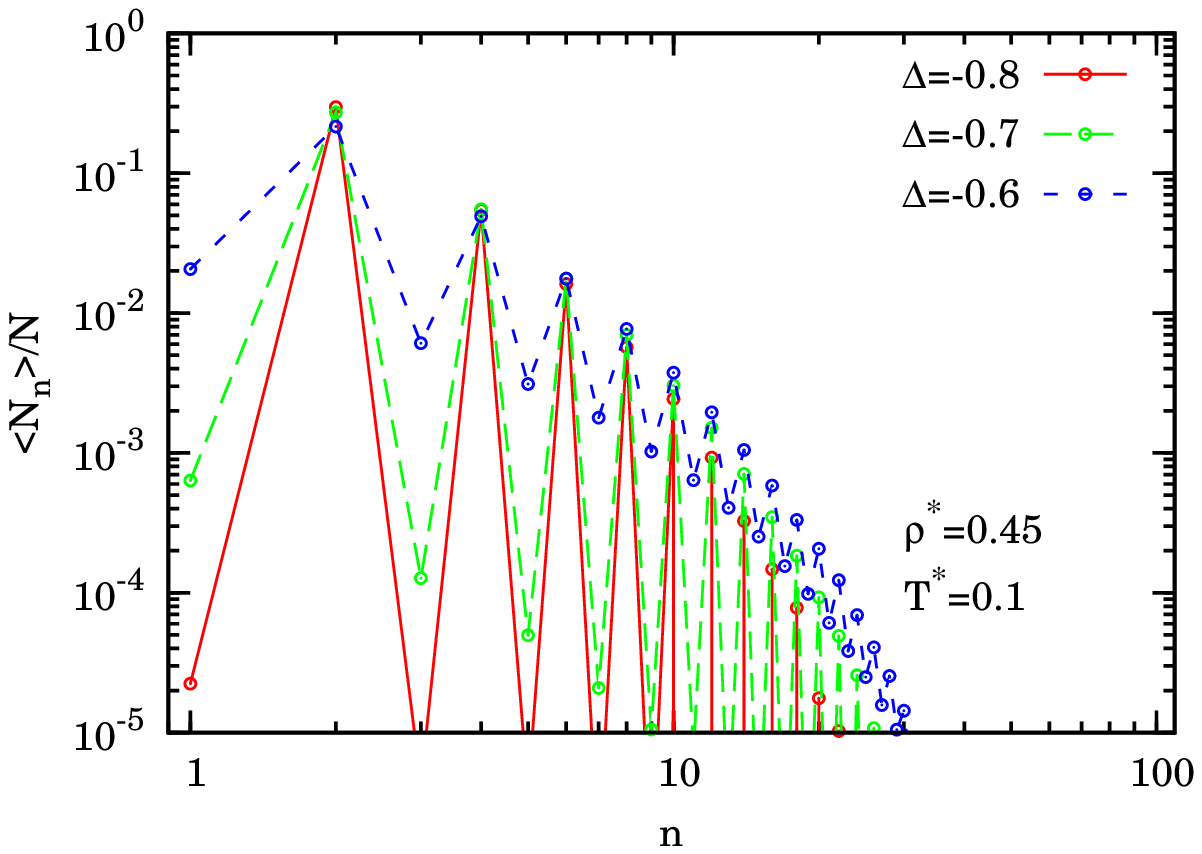}
\end{center}  
\caption{(Color online) Clustering properties of the fluid at $T^*=0.1$ and
$\rho^*=0.45$ at various values of negative nonadditivity approaching
$-1$. $N_n$ are the number of  
clusters made of $n$ particles. In the MC simulations we used $N=100$
particles and a number of MCS$=5\times 10^7$.}
\label{fig:dip}
\end{figure}

In Fig. \ref{fig:dip} we show the clustering analysis for the fluid
with $\Delta$ approaching $-1$ at $T^*=0.1$ and $\rho^*=0.45$. We see
how letting $\Delta$ approach $-1$ this stabilizes the neutrally charged
clusters and lowers the degree of dissociation. The first stable
cluster is the dipole: the ``overlap'' of a positive and a negative
sphere. These are dipoles of moment $qr_{12}$ with $r_{12}<\sigma
(1+\Delta+\delta^c)$ which may lack a gas-liquid criticality
\cite{Rovigatti2011}. We clearly have a transition from a conducting 
to an insulating phase as $\Delta$ goes from $0$ to $-1$.  

\begin{figure}[htbp]
\begin{center}
\includegraphics[width=8cm]{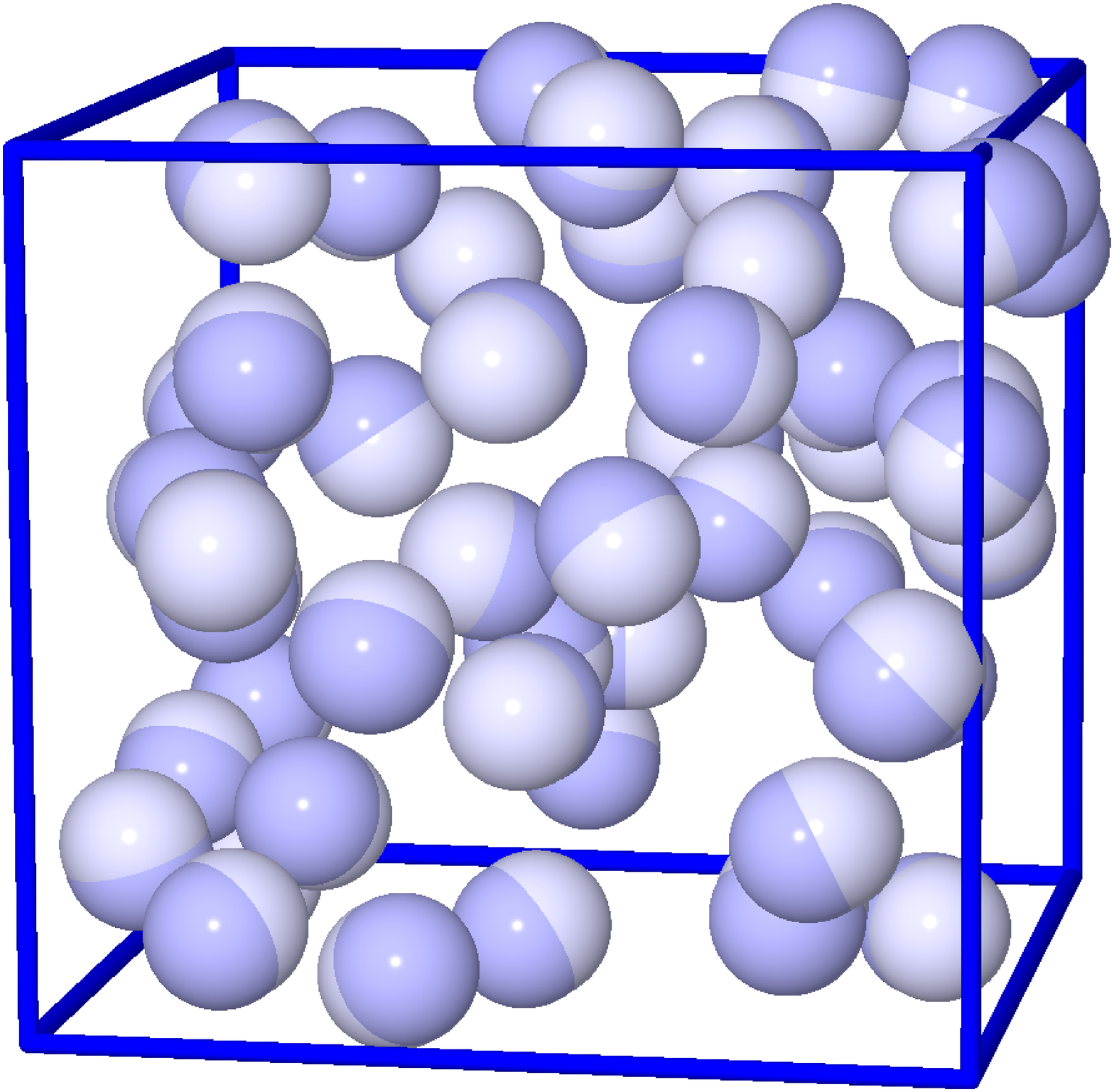}
\end{center}  
\caption{(Color online) Snapshot of the fluid at $T^*=0.1, \rho^*=0.45$ and
  $\Delta=-0.9$ showing the formation of the dipoles.}
\label{fig:snap-dip}
\end{figure}

In Fig. \ref{fig:snap-dip} we show a snapshot of the equilibrated
fluid at $T^*=0.1, \rho^*=0.45$ and $\Delta=-0.9$ from which one can
see the formation of the dipoles. We expect that in the limiting case
of $\Delta=-1$ the fluid we obtain is well reproduced by hard-spheres at
half the
density. This is confirmed by a comparison of the like radial
distribution functions with the one of the hard-spheres even if the
$\Delta=-1$ fluid simulation rapidly slows down 
into the frozen configuration of the overlapping anions and cations. In
order to overcome this problem one should alternate single particle
moves to cluster moves where one moves the center of mass of the
neutrally charged pairs. 

%

\subsection{Radial distribution function and structure factor}
\label{sec:rdf}

\begin{figure*}[htbp]
\begin{center}
\includegraphics[width=8cm]{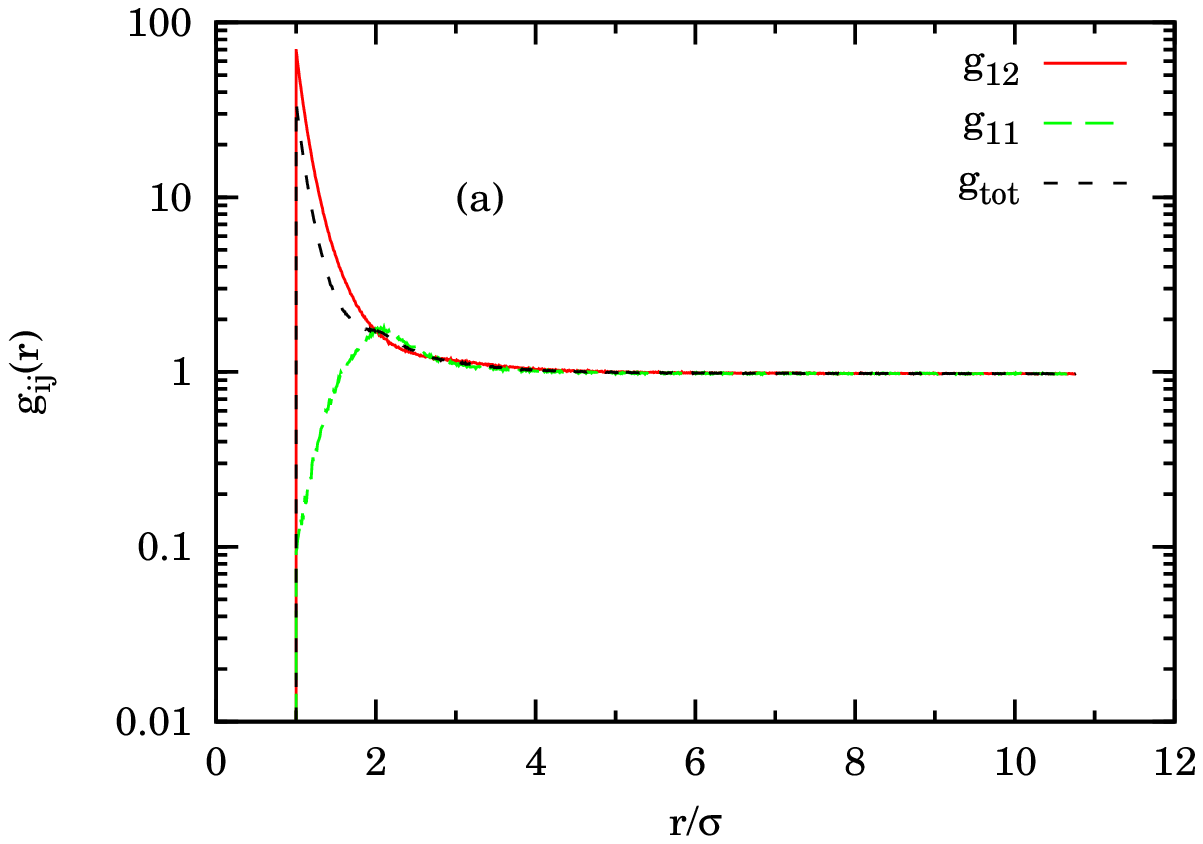}
\includegraphics[width=8cm]{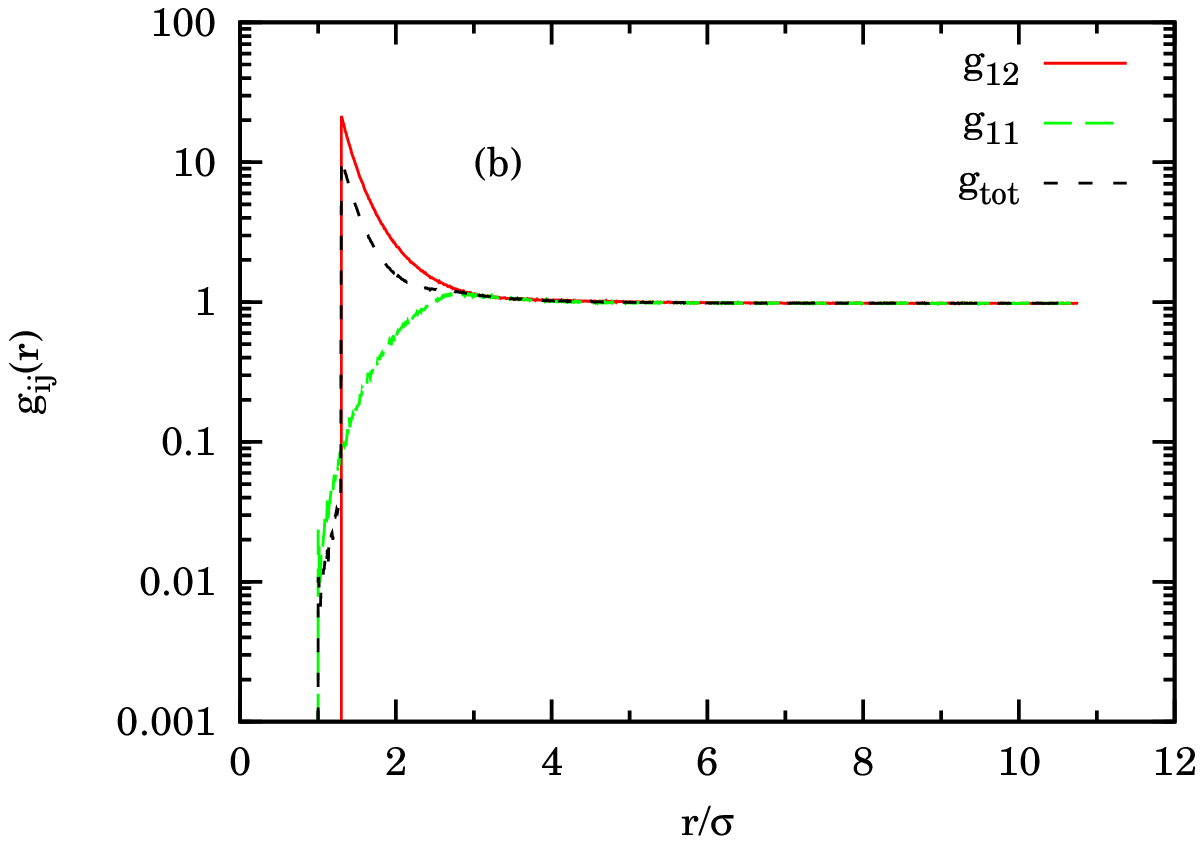}\\
\includegraphics[width=8cm]{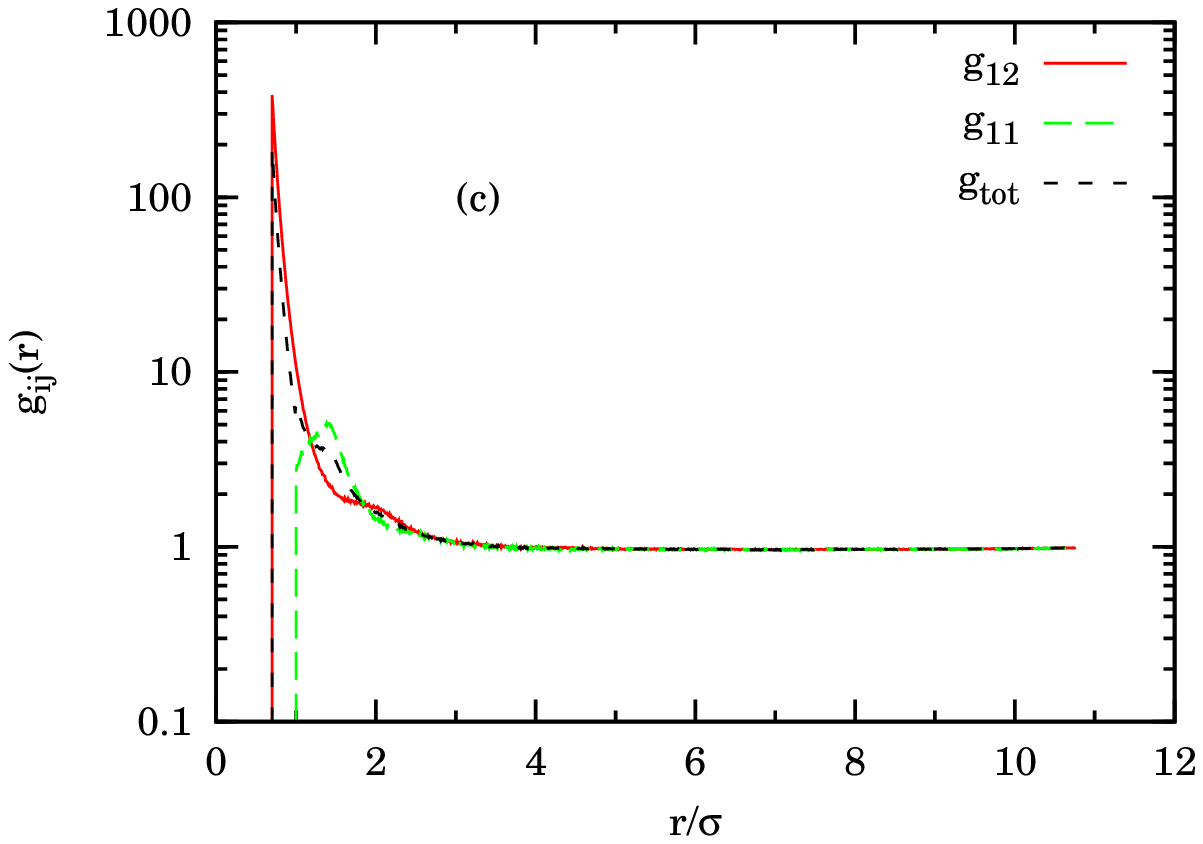}
\end{center}  
\caption{(Color online) Partial and total RDF in the
  simulations at $\rho^*=0.01, T^*=0.1,$ and
  $\Delta=0$ (panel (a)), $\Delta=0.3$ (panel (b)),
  $\Delta=-0.3$ (panel (c)). The reduced excess internal energy per
  particle of the fluid was in the three cases
  $U^\text{ex}/N=-0.3924(1), -0.29120(7), -0.6339(1)$ respectively.}
\label{fig:gr-r0.01-t0.1}
\end{figure*}

\begin{figure*}[htbp]
\begin{center}
\includegraphics[width=8cm]{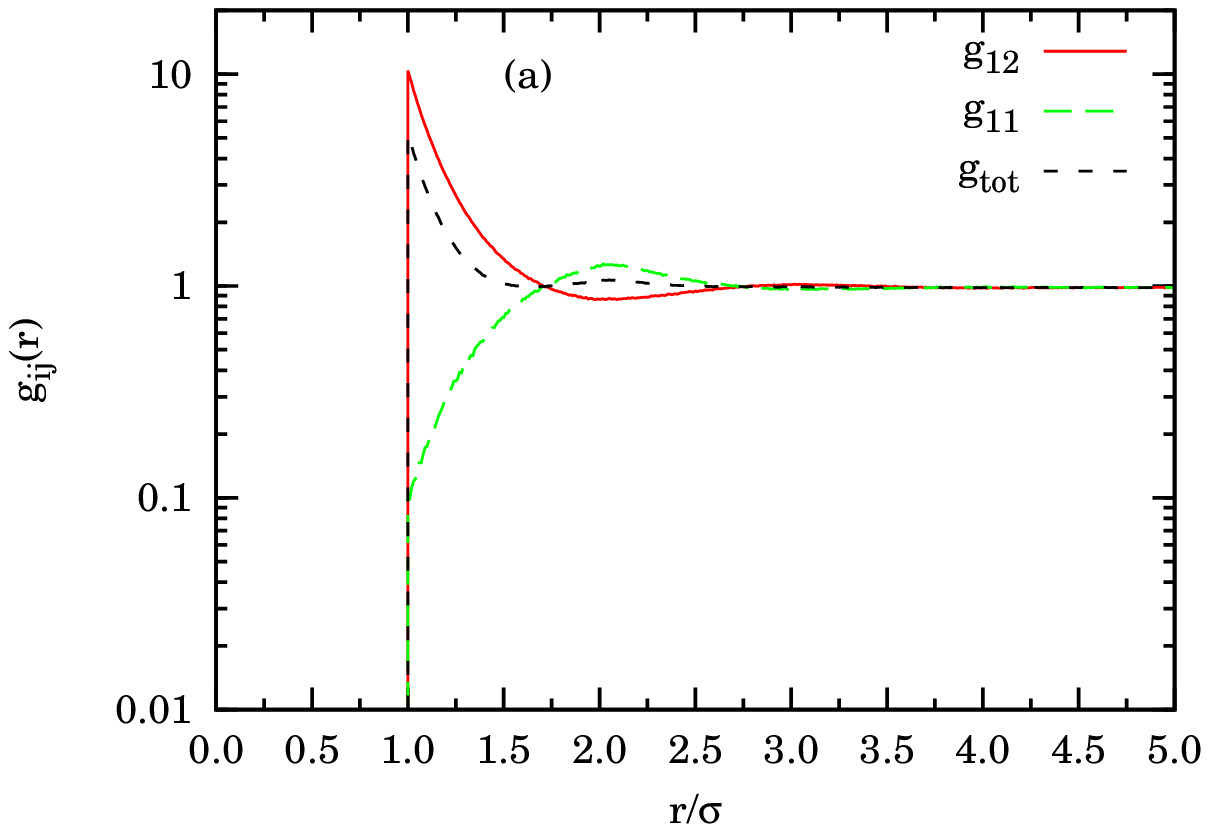}
\includegraphics[width=8cm]{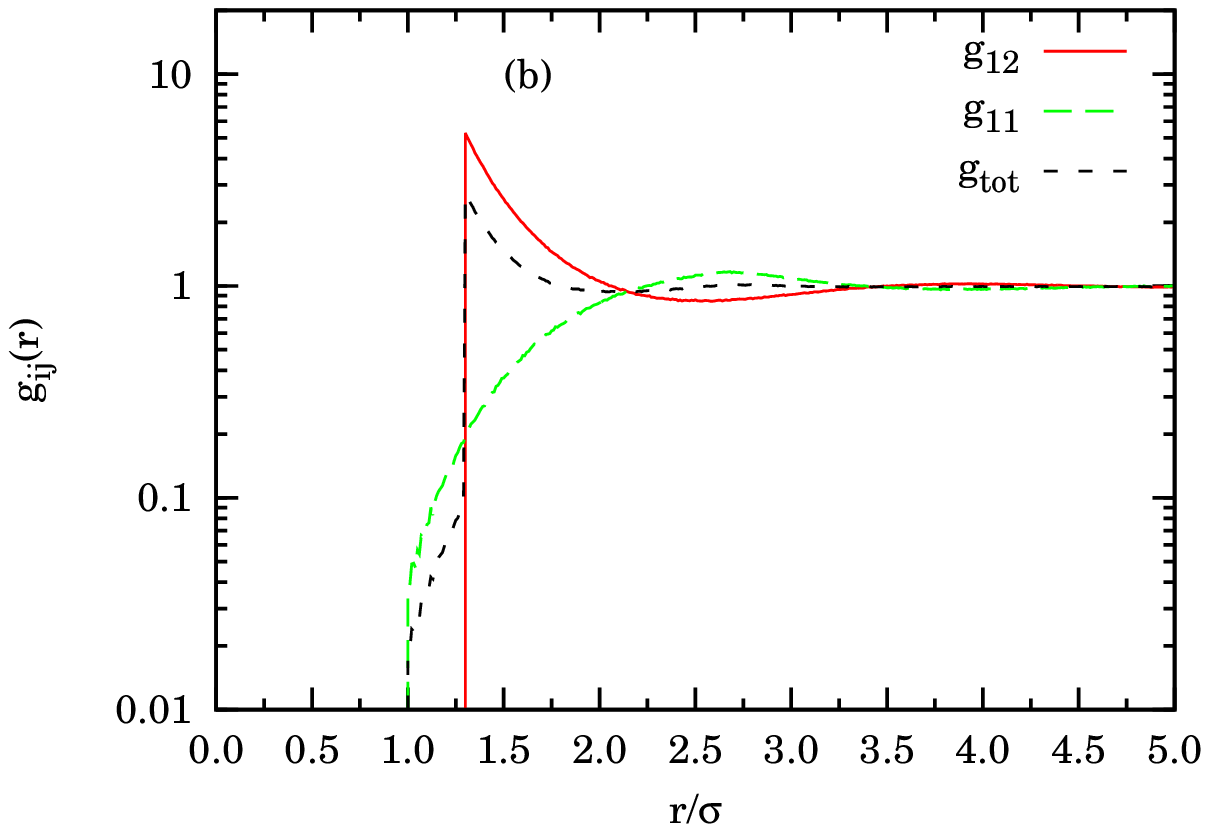}\\
\includegraphics[width=8cm]{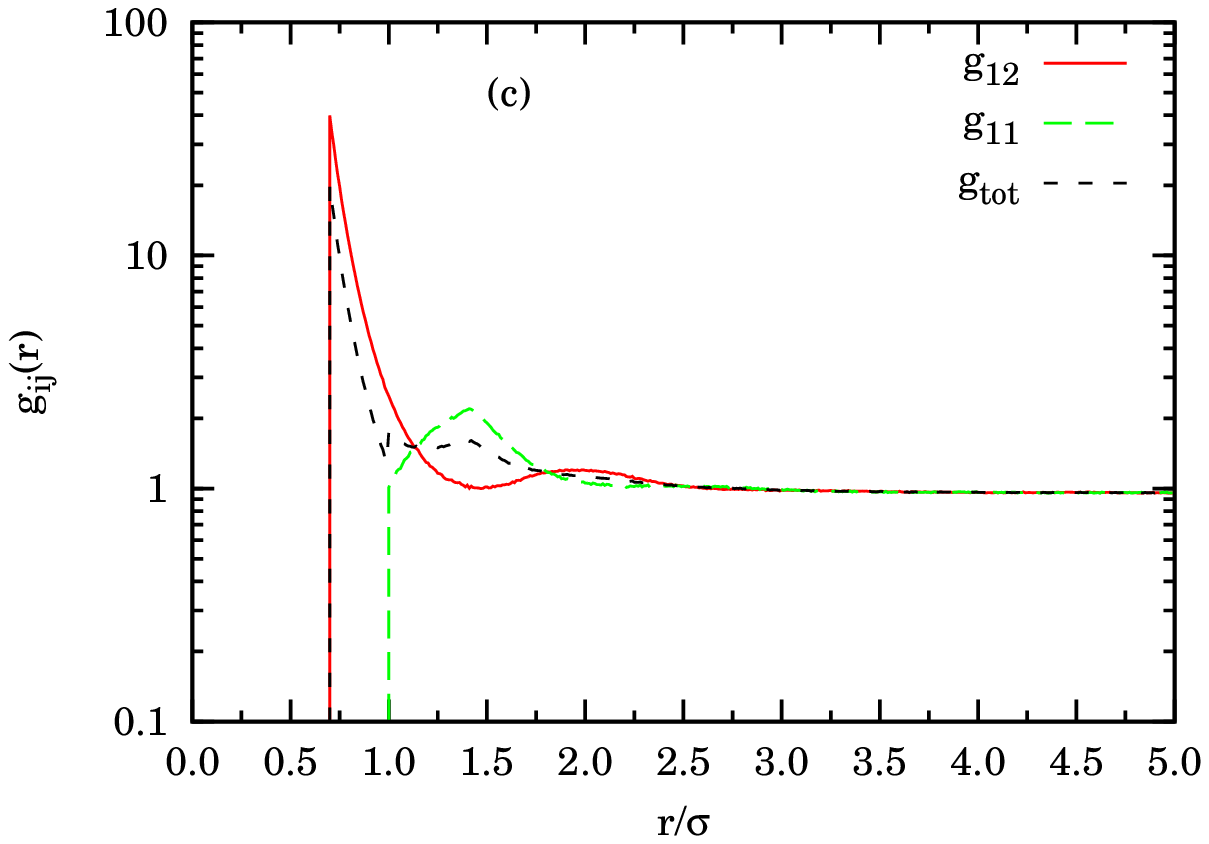}
\end{center}  
\caption{(Color online) Same as Fig. \ref{fig:gr-r0.01-t0.1} at $\rho^*=0.1$ and
  $T^*=0.1$. The reduced excess internal energy per
  particle of the fluid was in the three cases
  $U^\text{ex}/N=-0.50589(8), -0.41208(6), -0.7179(1)$ respectively.}
\label{fig:gr-r0.1-t0.1}
\end{figure*}

\begin{figure*}[htbp]
\begin{center}
\includegraphics[width=8cm]{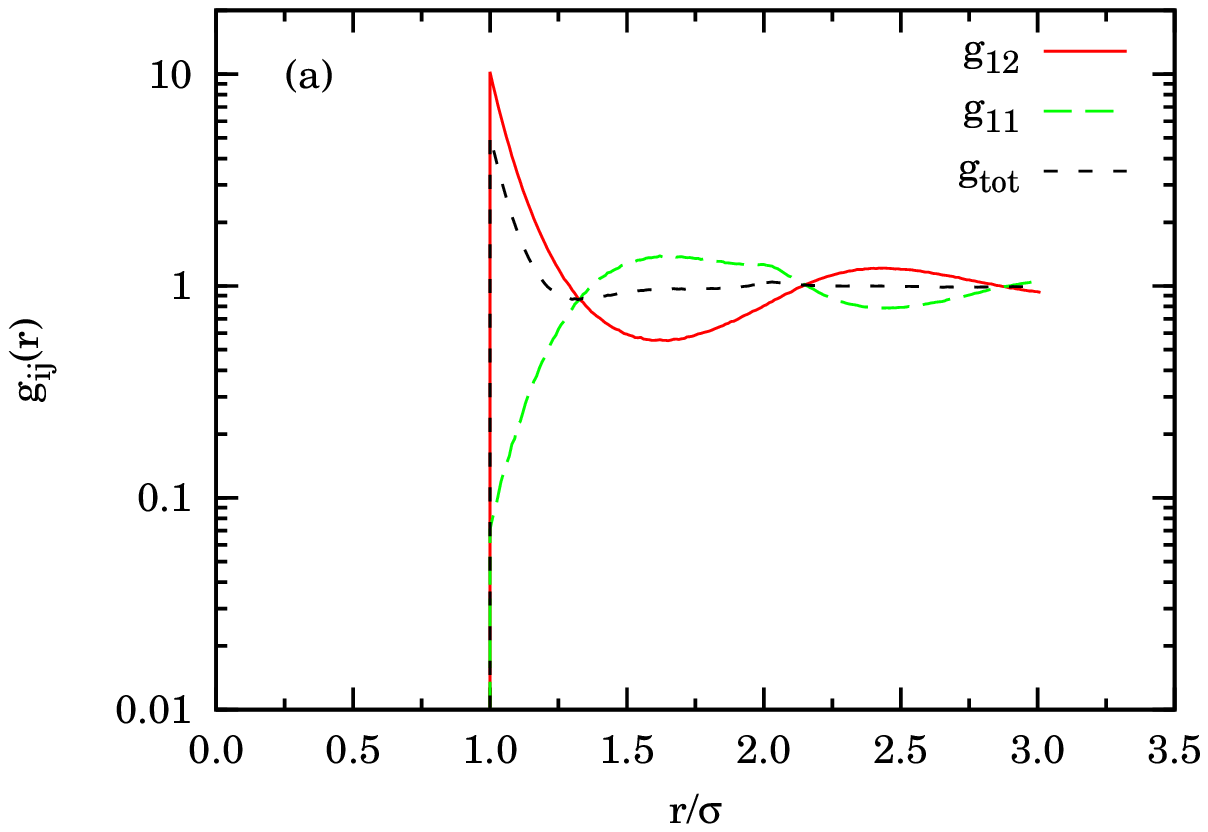}
\includegraphics[width=8cm]{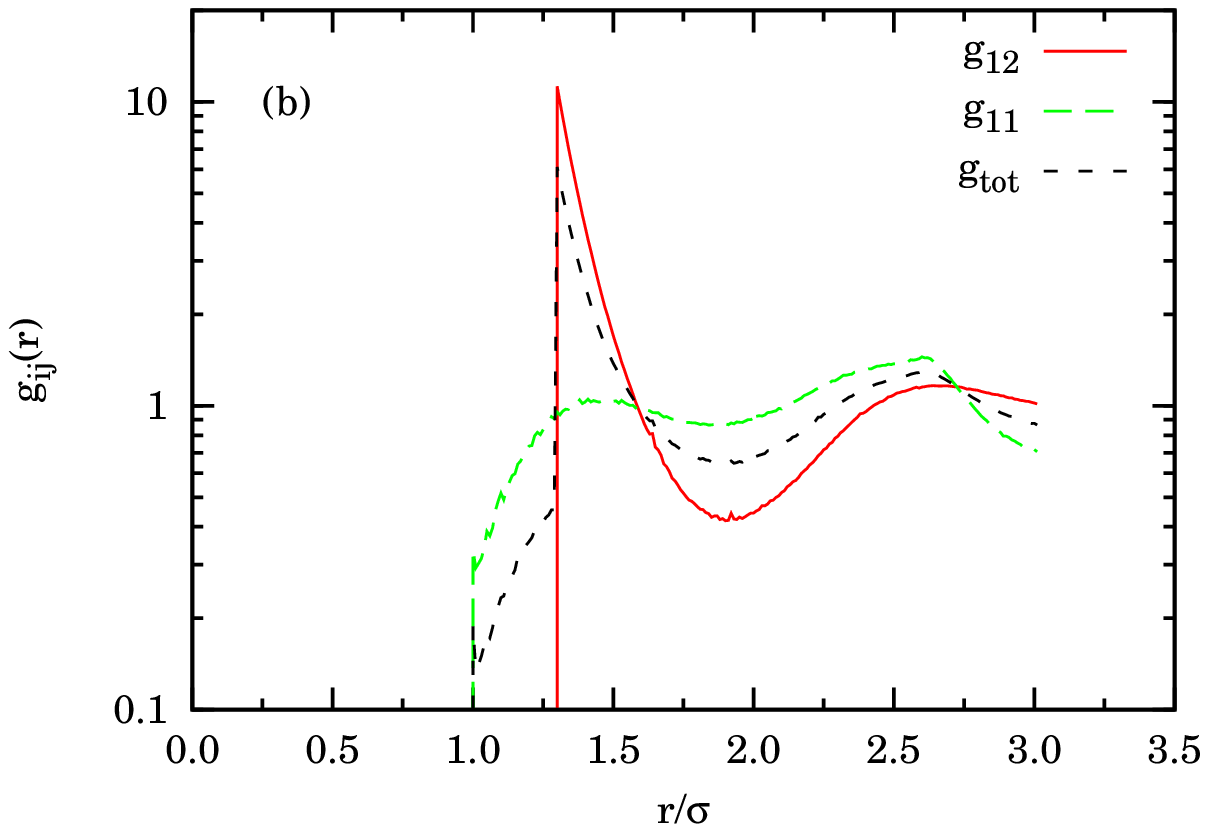}\\
\includegraphics[width=8cm]{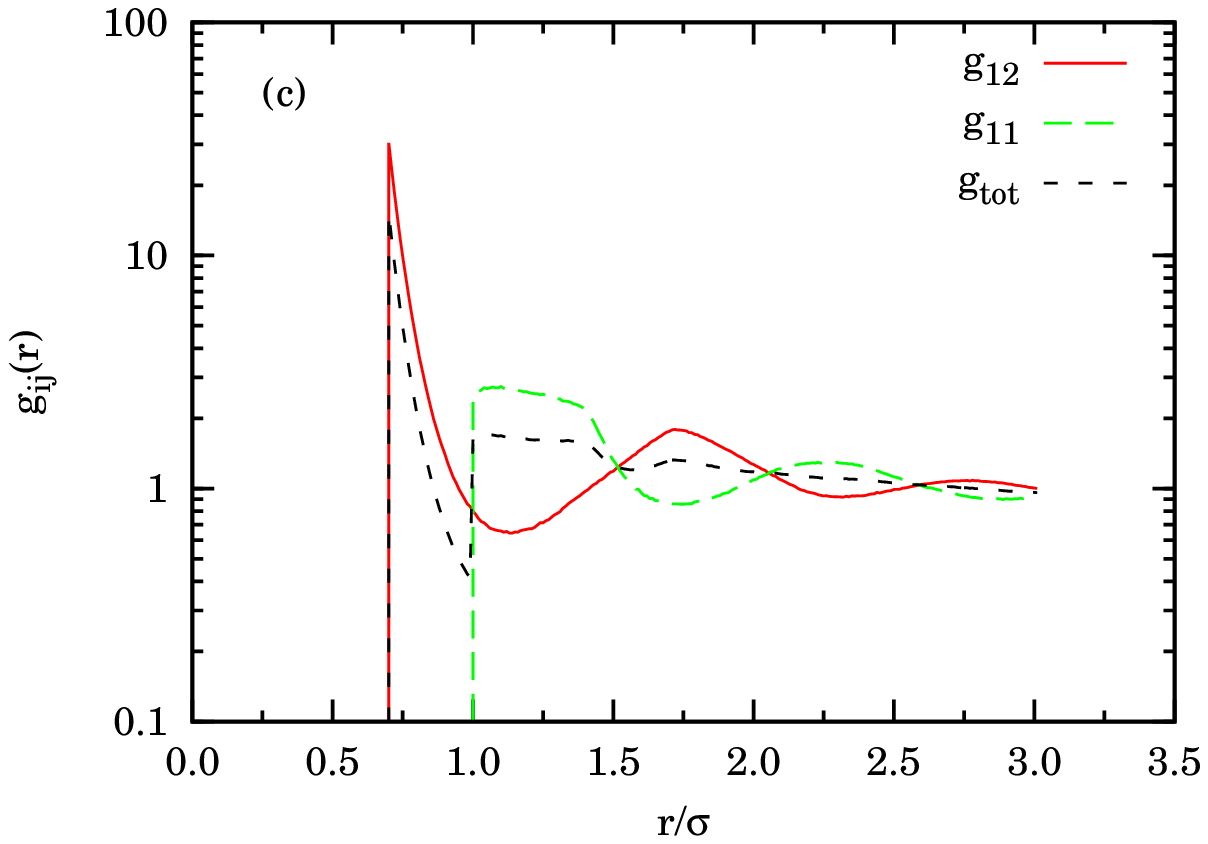}
\end{center}  
\caption{(Color online) Same as Fig. \ref{fig:gr-r0.01-t0.1} at $\rho^*=0.45$ and
  $T^*=0.04$. The reduced excess internal energy per
  particle of the fluid was in the three cases
  $U^\text{ex}/N=-0.69094(4), -0.55242(5), -0.96144(7)$ respectively.}
\label{fig:gr-r0.45-t0.04}
\end{figure*}

In Figs. \ref{fig:gr-r0.01-t0.1}-\ref{fig:gr-r0.45-t0.04} we show the
partial radial distribution functions (RDF),
$g_{ij}(r)=\langle\sum_{\mu\nu}^\prime\delta(\rr+\rr_\nu^j-\rr_\mu^i)\rangle
/N\rho x_ix_j$, where $\rr_\mu^i$ denote the position of particle
$\mu$ of species $i$ and the prime over the sum indicates that the
terms $\mu=\nu$ when $i=j$ are omitted, and the total
RDF, $g_{tot}=\sum_{i,j=1}^2g_{ij}x_ix_j$, of the three fluids
$\Delta=0,\pm 0.3$ 
at the thermodynamic states $T^*=0.1, \rho^*=0.01,0.1$ and
$T^*=0.04, \rho^*=0.45$. Of course, the restrictions $x_1=x_2$ and
$\sigma_{11}=\sigma_{22}$ imply $g_{11}=g_{22}$. In the simulations we
use $N=100$. 
 
From Fig. \ref{fig:gr-r0.01-t0.1} we see how the contact value of the
like RDF in the $\Delta=-0.3$ case is higher than in the additive
case and in the $\Delta=+0.3$ case is lower than in the additive
case. The contact value of the unlike RDF is highest for negative
nonadditivity indicating the tendency to form cation-anion pairs.

From Fig. \ref{fig:gr-r0.1-t0.1} we see again the same behaviors of
the contact values of the like and unlike RDF. In the negatively
additive case we begin to see an alternation of distribution of
oppositely charged shells of ions around a reference ion.

From Fig. \ref{fig:gr-r0.45-t0.04} we see how at this high density the
contact value of the like RDF is highest in the $\Delta=-0.3$ case but
in the $\Delta=+0.3$ case is still higher than in the additive
case. At $\Delta=-0.3$ we see clearly the formation of a 
second peak in the unlike RDF around $2+\Delta$ and the expected
alternation between the peaks of the like RDF with the ones of the
unlike RDF also present in the additive case. This alternation is not
present in the positively non-additive case indicating now the tendency
of like particles to cluster on a microscopic scale: like particles
penetrate inside the shell of unlike particles around a given
reference ion. The contact value of the unlike RDF is highest for
negative nonadditivity indicating the tendency to form cation-anion
pairs. 

\begin{figure*}[htbp]
\begin{center}
\includegraphics[width=8cm]{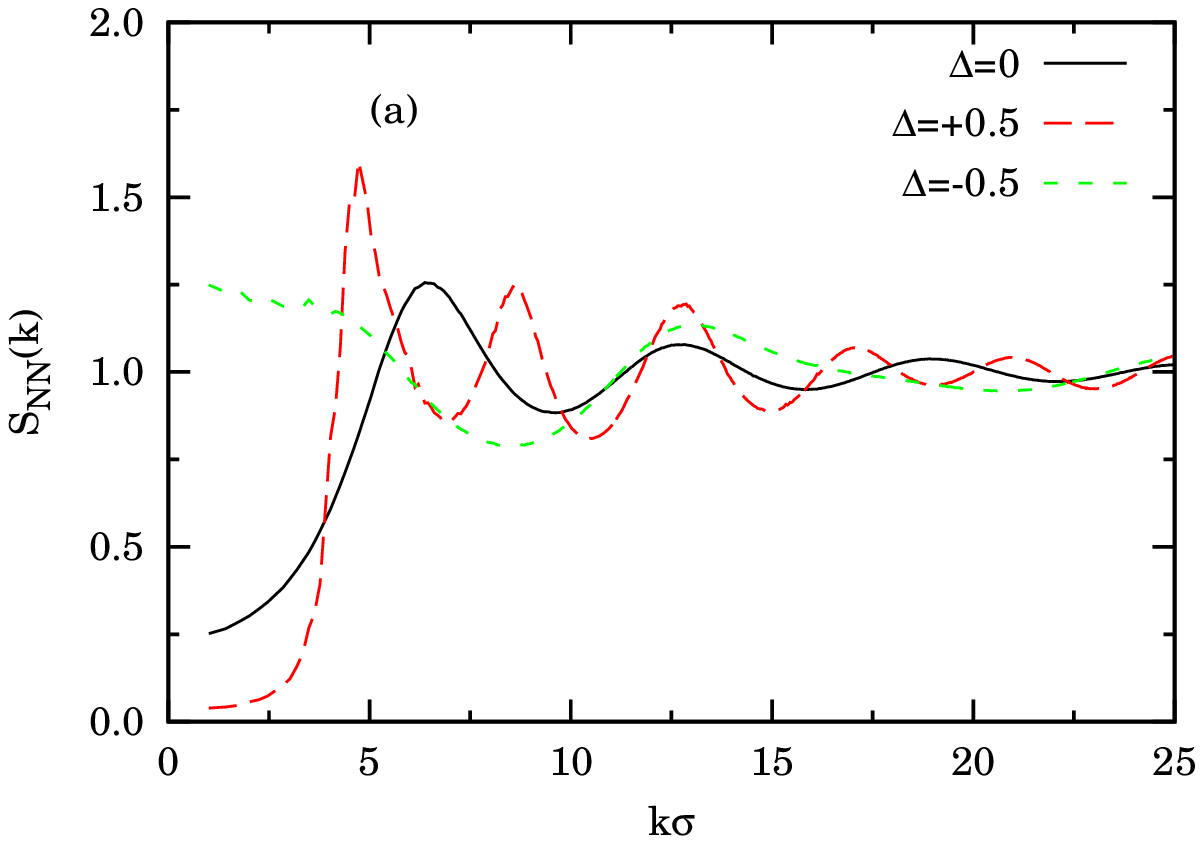}
\includegraphics[width=8cm]{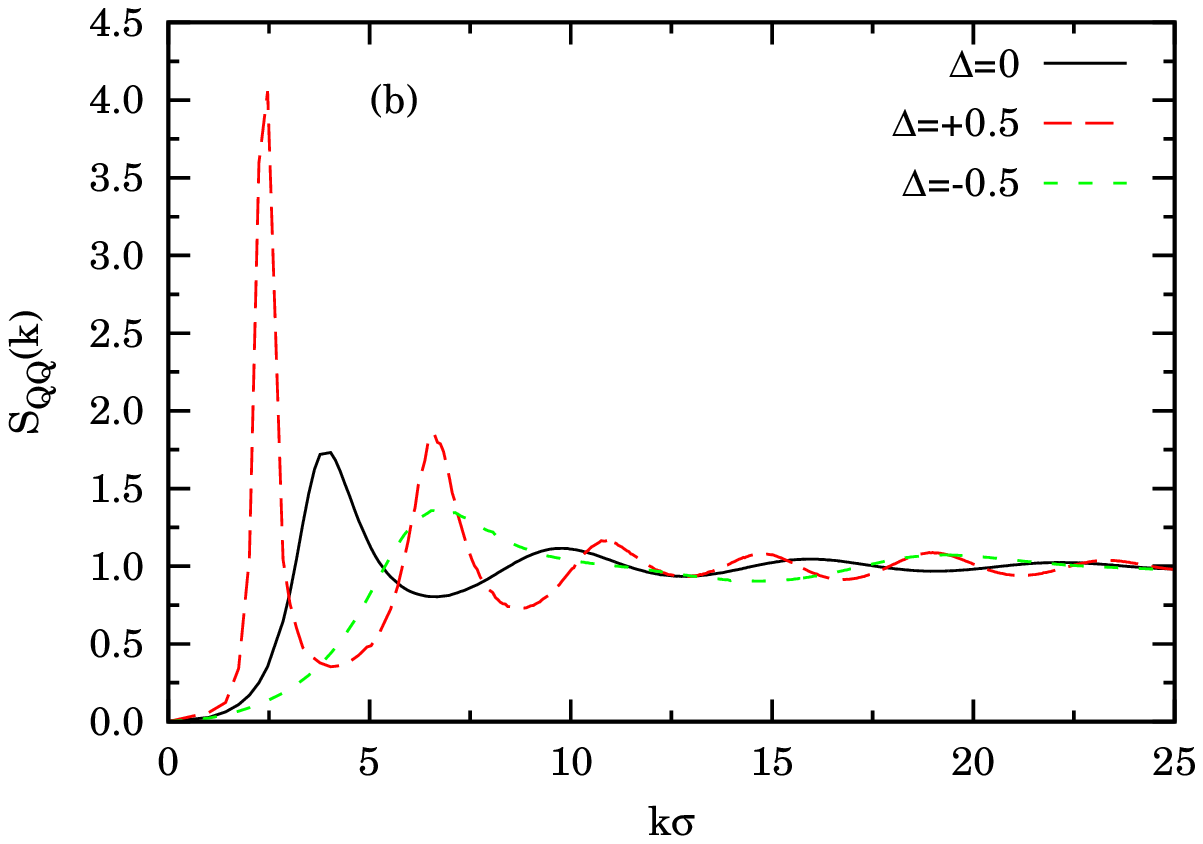}
\end{center}  
\caption{(Color online) Bhatia-Thornton structure factors $S_{NN}(k)$
  (panel (a)) and
  $S_{QQ}(k)$ (panel (b)) for $\rho^*=0.41253, T^*=0.12,$ and
  $\Delta=0,\pm0.5$ as in Fig. 1 of Ref. \cite{Pastore1985}. Note that
  our abscissa has to be divided by $1.2$ in order to compare with the
  units used in such reference.}
\label{fig:sk}
\end{figure*}

In Fig. \ref{fig:sk} we show the Bhatia-Thornton \cite{Bhatia1970}
structure factors $S_{NN}=[S_{11}+S_{22}+2S_{12}]/2$ and
$S_{QQ}=[S_{11}+S_{22}-2S_{12}]/2$ where
$S_{ij}(k)=\langle\rho^i_\kk\rho^j_{-\kk}\rangle/N\sqrt{x_ix_j}$ are
the partial structure factors and 
$\rho^i_\kk=\sum_\mu\exp(-i\kk\cdot\rr_\mu^i)$ is the Fourier
transform of the microscopic density of particles of species $i$.
In the figure we chose the same thermodynamic state and nonadditivity
considered in Fig. 1 of Ref. \cite{Pastore1985}. The positive
nonadditivity case has percolating clusters. From the figure we see that the
charge-charge structure factor $S_{QQ}$ tends to zero at $k=0$ a
consequence of electroneutrality in charged systems \cite{Hansen}
which suppresses long-wavelength fluctuations. In order to enforce
this condition the structure factor needs to develop a peak at 
small $k$ which reflects an essentially alternating distribution of
oppositely charged shell of ions around a reference ion. This type of
short range order is an indication of the tendency to cluster. From
the figure we see that at high density the positive non-additive fluid
tends to cluster more than the additive fluid and the negative
non-additive fluid tends to cluster less than the additive fluid, in
agreement with the results presented in the previous Section. With
regard to the number-number structure factor $S_{NN}$ we see that as
the nonadditivity decreases the isothermal compressibility $S_{NN}(0)$
(see the appendix of Ref. \cite{Bhatia1970,Fantoni05b}) increases and
the short range order is reduced. 

In Table \ref{tab:thermo} we report the excess internal energy per particle
$U^\text{ex}/N=\epsilon\sigma \langle {\cal U}\rangle/Nq^2$, the
compressibility factor $Z=\beta P/\rho$, and the total clusters
concentration $\sum_n\langle N_n\rangle/N$ for the cases simulated. The
compressibility factor is calculated according to the virial theorem
\bq \nonumber
Z&=&1+\frac{U^\text{ex}}{3NT^*}\\
&&+\frac{\pi\rho^*}{3}[g_{11}(\sigma)+(1+\Delta)^3g_{12}(\sigma(1+\Delta))]~.
\eq
If the clusters do not interact, as in the independent cluster model
(ICM) of Gillan \cite{Gillan1983}, one should have
$Z^\text{ICM}=\sum_n\langle N_n\rangle/N$. From Table \ref{tab:thermo}
we can see how this condition is never satisfied in the cases
considered.  
\begin{table}[htbp]
\caption{Excess internal energy per particle
  $U^\text{ex}$ $=$ $\epsilon\sigma \langle {\cal U}\rangle/q^2$,
  compressibility factor $Z=\beta P/\rho$, and total clusters
  concentration $\sum_n\langle N_n\rangle/N$ for the cases simulated.}  
\label{tab:thermo}
{\scriptsize
\begin{center}
\begin{tabular}{|c|c|c||c|c|c|}
\hline
$T^*$ & $\rho^*$ & $\Delta$ & $-U^\text{ex}/N$ & $Z-U^\text{ex}/3NT^*$
& $\sum_n\langle N_n\rangle/N$ \\ 
\hline
0.1 &0.45&$0   $& 0.62711(9)& 3.764(5) & 0.317  \\
0.1 &0.45&$+0.3$& 0.46212(9)& 9.16(1)  & 0.026 \\
0.1 &0.45&$-0.3$& 0.81357(9)& 3.019(3) & 0.410 \\
\hline
0.1 &0.3 &$0   $& 0.58827(6)& 2.869(3) & 0.528 \\
0.1 &0.3 &$+0.3$& 0.47493(7)& 4.837(6) & 0.255 \\
0.1 &0.3 &$-0.3$& 0.7814(1) & 2.797(3) & 0.483 \\
\hline
0.1 &0.2 &$0   $& 0.55390(6)& 2.445(2) & 0.637 \\
0.1 &0.2 &$+0.3$& 0.45639(6)& 3.231(3) & 0.540 \\
0.1 &0.2 &$-0.3$& 0.75483(9)& 2.657(3) & 0.530 \\
\hline
0.1 &0.1 &$0   $& 0.50589(8)& 2.098(2) & 0.730 \\
0.1 &0.1 &$+0.3$& 0.41208(6)& 2.218(2) & 0.747 \\
0.1 &0.1 &$-0.3$& 0.7179(1) & 2.539(3) & 0.579 \\
\hline
0.1 &0.01&$0   $& 0.3924(1) & 1.7373(8)& 0.830 \\
0.1 &0.01&$+0.3$& 0.29120(7)& 1.493(3) & 0.900 \\
0.1 &0.01&$-0.3$& 0.6339(1) & 2.409(2) & 0.652 \\ 
\hline
0.1 &0.001&$0   $& 0.3076(1) & 1.582(1) & 0.870 \\
0.1 &0.001&$+0.3$& 0.1971(1) & 1.2962(6)& 0.943 \\
0.1 &0.001&$-0.3$& 0.5992(1) & 2.355(2) & 0.677 \\ 
\hline
0.04&0.45&$0   $& 0.69094(4)& 5.863(8) & 0.104 \\
0.04&0.45&$+0.3$& 0.55242(5)& 12.83(2) & 0.012 \\
0.04&0.45&$-0.3$& 0.96144(7)& 7.09(1)  & 0.028 \\ 
\hline
0.04&$5\times 10^{-5}$&$0   $& 0.48804(2)& 4.112(1) & 0.563 \\
0.04&$5\times 10^{-5}$&$+0.3$& 0.35342(2)& 3.254(1) & 0.681 \\
0.04&$5\times 10^{-5}$&$-0.3$& 0.69764(1)& 5.230(2) & 0.493 \\ 
\hline
\end{tabular}
\end{center}
}
\end{table}
%

\subsection{Gas-liquid coexistence} 
\label{sec:binodal}

An important question we try to answer is how the gas-liquid
coexistence curve of the pure RPM fluid changes upon switching on of the
nonadditivity parameter. To this aim we first perform a density
distribution analysis within the $NVT$ ensemble which allows us to
easily extract a semi-quantitative result and then we use the Gibbs
ensemble technique for a careful quantitative determination of the
binodals.    

\subsubsection{The density distribution approach} 
\label{sec:distribution}

Sufficiently close to the critical point we determine how
semi-quantitatively the behavior of the gas-liquid coexistence region changes 
by switching on a negative or a positive nonadditivity. To this
aim we divide the simulation box into $m^3$ cubes of side $L_c=L/m$ and
register, as the run progresses, the density inside each cell
$\rho_i={\cal N}_i/L_c^3$, where ${\cal N}_i$ is the number of
particles inside the $i$th cell so that $\sum_{i=1}^{m^3}{\cal
  N}_i=N$. Then we calculate the density 
distribution function \cite{Rovere1988,*Rovere1990,*Rovere1993}
$P_m(\rho)=\sum_{i=1}^{m^3}P_m(\rho_i)/m^3$, where $P_m(\rho_i)$ is
the distribution function for the $i$th cell, with $\int P_m(\rho)\,
d\rho=1$. Above 
the critical temperature the density probability distribution function
can be described by a Gaussian centered at the simulation density
whereas below it becomes bimodal with two peaks one centered at the
gas density and one at the liquid density.

We start from an initial configuration of particles of random species
placed on a simple cubic lattice. We equilibrate (melt) the fluid for
$10^6$ MCS$/$particle. We then sample the distribution function
every $10$ MCS. To allow the particles to diffuse out of the
cells we choose the subdivision of the simulation box in cells with a
random displacement $\rr=(r_x,r_y,r_z)$ with $r_x,r_y,r_z\in
[0,L]$. This procedure turned out to greatly enhance the efficiency
of the determination of the cell density distribution. And we measure
the distribution function on runs of $1\times 10^6$ MCS$/$particle.  

\begin{figure*}[htbp]
\begin{center}
\includegraphics[width=8cm]{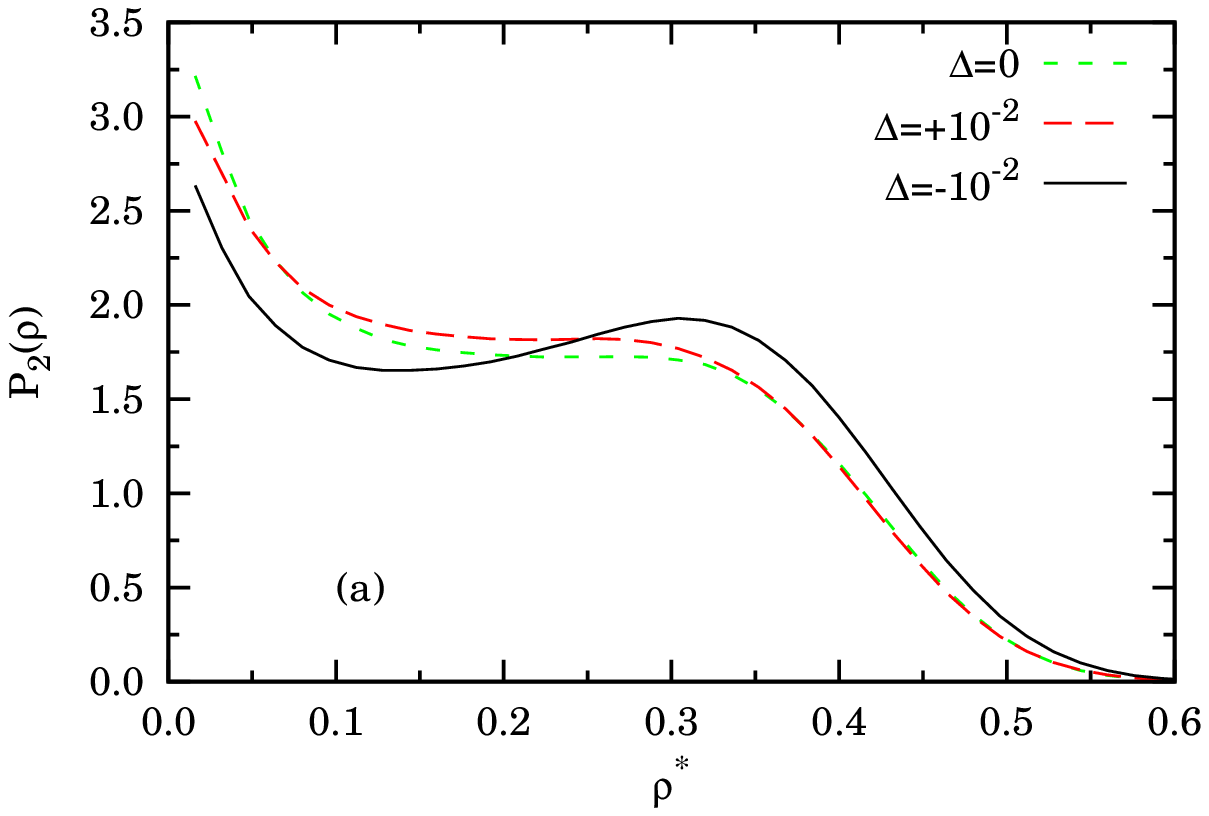}
\includegraphics[width=8cm]{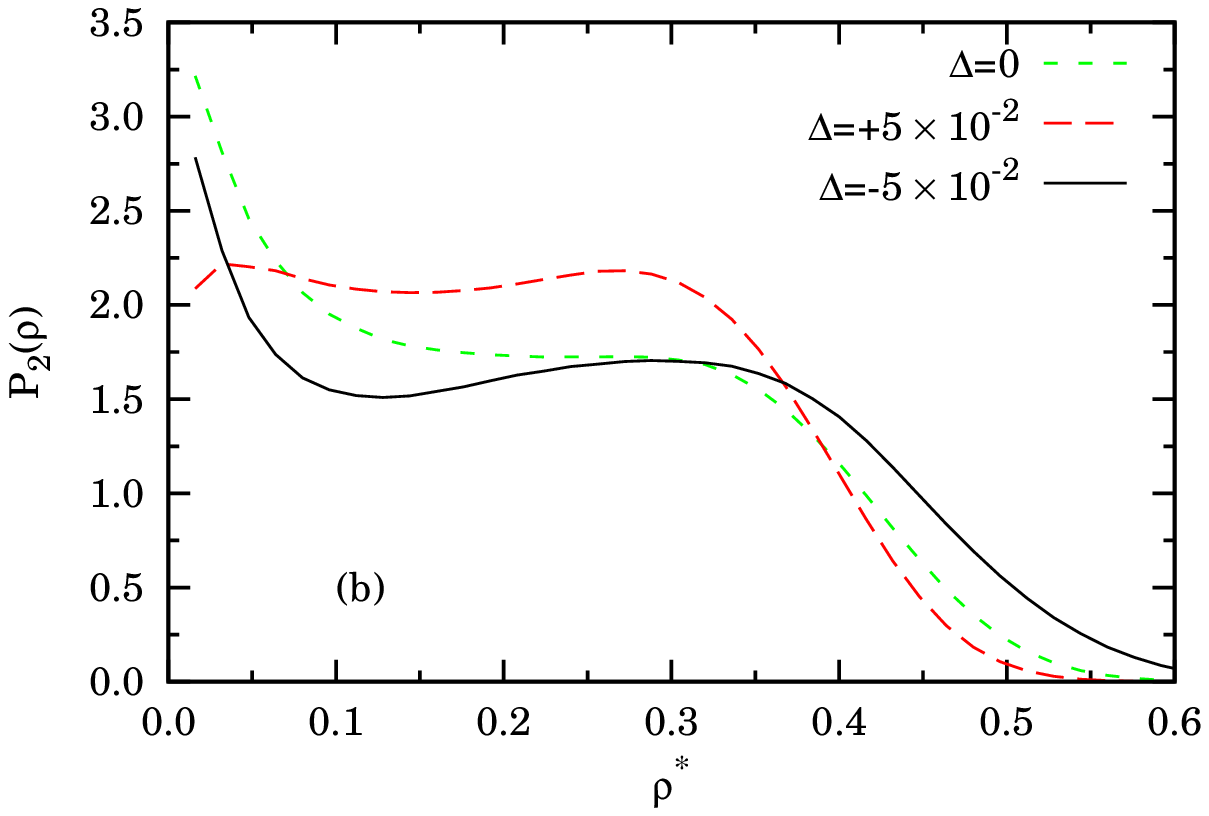}\\
\includegraphics[width=8cm]{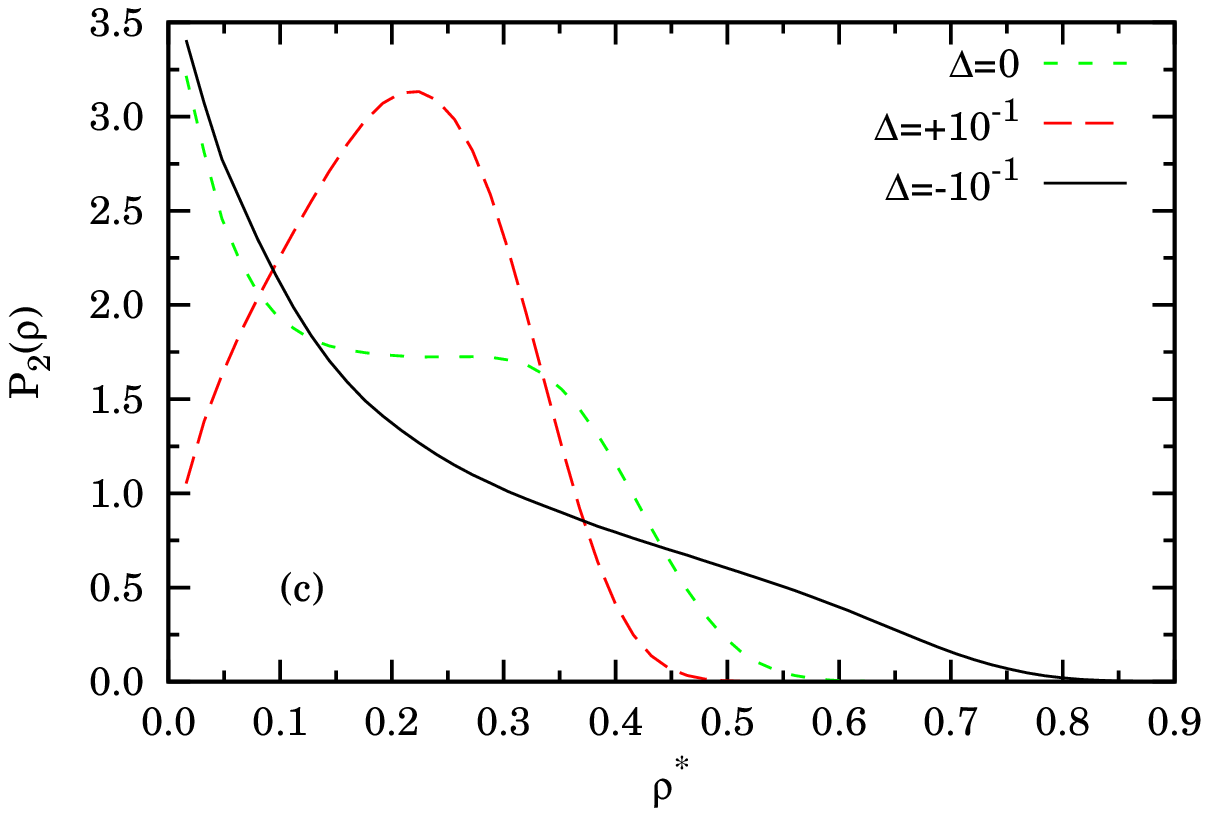}
\end{center}  
\caption{(Color online) Cell density distribution function for the fluid at
  $T^*=0.025, \rho^*=0.2$ and $\Delta=0,\pm{\cal D}$ with
  ${\cal D}=10^{-2},5\times 10^{-2},10^{-1}$ in panels (a), (b), and
  (c) respectively. We used $N=100$
  and $m=2$ with $1\times 10^6$ MCS$/$particle.}   
\label{fig:bim-r0.2-t0.02}
\end{figure*}

Choosing $m=2$ and $N=100$ the results for the fluid at a temperature
$T^*=0.025$ above the triple point of the RPM \cite{Vega2003}, a
density $\rho^*=0.2$ well within the coexistence region of the pure 
RPM fluid, and $\Delta=0,\pm {\cal D}$ with ${\cal D}=10^{-1}, 10^{-2},
5\times 10^{-2}$ are shown in Fig. \ref{fig:bim-r0.2-t0.02}. In this
case the minimum density that can be registered is $1/L_c^3=0.2\times
8/100=0.016$. We see that the pure RPM fluid shows a density
distribution function with two peaks: The first one, which
lies below the minimum density (and is not visible in our data), at
approximately the low density of the gas phase and the second one at
approximately the high density of the liquid phase around a reduced
density of $0.3$. At ${\cal D}=10^{-2}$ the positions of the peaks 
are roughly the same to the pure RPM. At ${\cal D}=5\times 10^{-2}$
the density of the liquid peak in the negatively non-additive fluid is
higher than the one of the pure RPM whereas the positively non-additive
fluid has a gas peak, now visible, at higher density than the pure RPM
and a liquid peak at lower density than the pure RPM. At ${\cal D}= 
10^{-1}$ this separation tends to increase: In the positively non-additive
model the critical temperature is too close to $0.025$ and the bimodal
is degenerate into a curve with a single peak centered on the simulation
density $0.2$ whereas in the negatively non-additive fluid the liquid
peak is changed into a broad tail extending up to a density of
$0.8$. This findings suggest that at a given temperature the width of
the coexistence region, relative to the one of the pure RPM, tends to
increase for the negatively non-additive model and to decrease for the
positively non-additive model. This result is made more clear and
precise in the following Section where we present our Gibbs Ensemble
Monte Carlo calculation.  
%

\subsubsection{Gibbs ensemble analysis}
\label{sec:gemc}

In order to quantitatively determine the gas-liquid coexistence line
of our fluid we use the Gibbs ensemble MC (GEMC) technique
\cite{Frenkel-Smit,*Panagiotopoulos87,*Panagiotopoulos88,*Smit89a,*Smit89b},
starting from the pure RPM one and gradually switching on the
nonadditivity. Here, we are not interested in the behavior really close
to the critical point but rather in the shape of the binodal curve and
how it moves as a function of $\Delta$.

The GEMC  method of Panagiotopoulos is
now widely adopted as a standard method for calculating
phase equilibria from molecular simulations. According to
this method, the simulation is performed in two boxes containing
the coexisting phases. Equilibration in each phase is
guaranteed by moving particles. Equality of pressures is satisfied
in a statistical sense by expanding the volume of one of
the boxes and contracting the volume of the other. Chemical
potentials are equalized by transferring particles from one
box to the other. Like the first simulations for the RPM performed by
Panagiotopoulos \cite{Panagiotopoulos92} we use single ion transfer
by introducing a background charge density to assure charge neutrality
at all times during the run. This way, the system remains overall
neutral, but the modified model is similar to a two component plasma 
and in a strict sense different from the original RPM model which
assumes a zero charge density for the background. To overcome the
electroneutrality problem Orkoulas {\sl et al.} \cite{Orkoulas1994}
considered pair transfers.

In the GEMC run we have at each step a probability $a_p/(a_p+a_v+a_s)$
for a particle random displacement, $a_v/(a_p+a_v+a_s)$ for a volume
change, and $a_s/(a_p+a_v+a_s)$ for a particle swap move between the
gas and the liquid box. We generally choose $a_p=1, a_v=1/10,$ and
$a_s=1$. The maximum particle displacement is kept equal to $L_i/1000$ where
$L_i$ is the side of the $i$th box with $i=1,2$. Regarding the volume
changes, following Ref. \cite{Frenkel-Smit} we perform a random walk
in $\ln[V_1/V_2]$, with $V_i$ the volume of the $i$th box choosing a
maximum volume displacement of $1\%-10\%$. Volume moves are
computationally the cheapest since the energy scales with the length
of the box with inverse proportionality. We generally use a total
number of $N=100$ particles except close to the critical point where it
proves necessary to increase the number of particles in order to avoid
large fluctuations in the two densities. We use $10-40$ million MCS
for the equilibration and $100-200$ million MCS for the production.

\begin{figure}[htbp]
\begin{center}
\includegraphics[width=10cm]{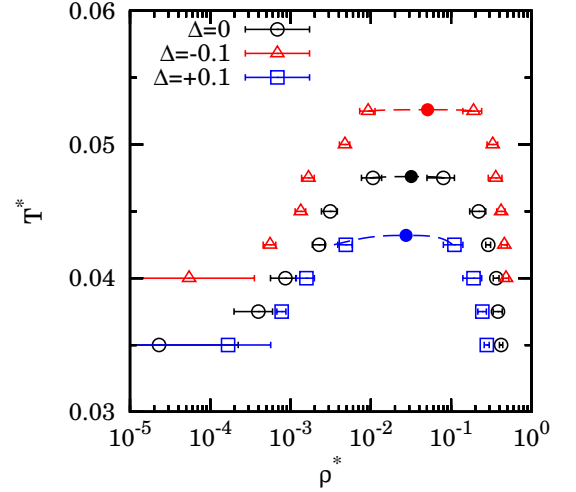}
\end{center}  
\caption{(Color online) Binodals obtained with the GEMC simulations. The dashed lines
  are the result of the extrapolation described in the text. The
  filled circles are the critical points.} 
\label{fig:gemc}
\end{figure}

The results are summarized in Table \ref{tab:gemc} and
Fig. \ref{fig:gemc}. Note that since we get the same coexistence
curve of Orkoulas {\sl et al.} \cite{Orkoulas1994} for the pure RPM,
as Fig. \ref{fig:coex} clearly shows, we consider as equivalent, 
at the present level of accuracy, our procedure, 
employing single neutralized particle
transfers, and the one of Orkoulas, where pair particle transfers
between the two boxes are used. This  can be 
justified by observing that
the fluctuations of charge in the various statistical physics
ensembles are expected to decay to zero with the system size and we
empirically find that in our case they are already practically
irrelevant. The only relevant difference we observe with respect to
the calculation of Orkoulas is the fact that in our case there is a
much more considerable emptying of the gas box at low temperatures
which may have some effect in the point at the lowest temperature. 
We do not carry 
out a systematic study of possible system size dependence of the
results but, for the pure RPM, we repeat the calculation at
$T^*=0.045$ and $0.0475$ for two different system sizes with the
largest being $N=370$. The comparison suggests that the critical point
tends to shift slightly at higher temperatures upon a system size
increase but far away from the critical point the coexistence curve is
not affected appreciably by the system size. However, we stress that
an accurate study of critical properties of the present model is beyond the 
scope of this work.

From the figure we can
see clearly the trend:  a positive nonadditivity tends to lower the critical
temperature whereas a negative one tends to push the binodal to higher
temperatures. This is in agreement with the findings from the density
distribution analysis previously presented. It is well known that RPM
condensation is almost identical to that of charged hard dumbbells,
underlining the fact that the vapor is essentially already fully
associated into dimers and higher neutral clusters, and that the liquid
structure and thermodynamics are only weakly perturbed by fusing ions
together. Hence, if one imagines cooling down on the critical isochore,
we can say that the critical point is reached  
when ion association is complete and 
then it becomes convenient for the system to phase separate. With positive
nonadditivity, ion association is less favorable and the critical
temperature must go down (association is complete only at lower
temperatures); with negative nonadditivity, ion association is more
favorable, and the critical temperature must go up (association is
complete already at higher temperatures).

\begin{table*}[htbp]
\caption{Phase coexistence properties for the pure RPM ($\Delta=0$)
  and the non-additive RPM ($\Delta\neq 0$). $T^*$ is the reduced
  temperature, $N$ is the total number of
particles in the system for a certain run, $N_g$ is the average number of
particles in the gas box during the run,
$\mu^*_l=\mu_l\epsilon\sigma/q^2-T^*\ln\Lambda^3$ is the reduced
chemical potential of the liquid box ($\Lambda$ being the de Broglie
thermal wavelength), $U^\text{ex}_i$ is the total excess internal
energy, and $\rho_i^*$ is the reduced density, of the gas phase 
$i=g$ and the liquid phase $i=l$.} 
\label{tab:gemc}
{\scriptsize
\begin{center}
\begin{tabular}{|c|c|c||c|c||c|c||c|c|}
\hline
$\Delta$ & $T^*$ & $N$ & $N_g/N$ & $-\mu_l^*$ & $-U^\text{ex}_g/N$ & $-U^\text{ex}_l/N$ &$\rho_g^*$ & $\rho_l^*$ \\ 
\hline
0 & 0.0475 & 370 & 0.51(1)  & 0.63(1)&0.547(1)& 0.609(1)  & $5.2(9)\times 10^{-3}$&0.11(3)\\
0 & 0.0475 & 200 & 0.33(1)  & 0.63(1)&0.559(2)& 0.604(1)  & $1.1(3)\times 10^{-2}$&0.08(3)\\
0 & 0.045  & 370 & 0.26(1)  & 0.69(3)&0.528(4)& 0.6400(7) & $2.3(5)\times 10^{-3}$& 0.22(5)\\
0 & 0.045  & 100 & 0.27(1)  & 0.63(2)&0.537(4)& 0.6393(9) & $3.1(7)\times 10^{-3}$& 0.22(5)\\
0 & 0.0425 & 100 & 0.166(8) & 0.65(1)&0.52(1) & 0.6576(8) & $2.3(4)\times 10^{-3}$& 0.29(2)\\
0 & 0.04   & 100 & 0.069(5) & 0.73(1)&0.50(2) & 0.6745(5) & $8(3)\times 10^{-4}$& 0.35(3)\\
0 & 0.0375 & 100 & 0.036(2) & 0.72(1)&0.4(1)  & 0.6835(5) & $4(2)\times 10^{-4}$& 0.38(5)\\
0 & 0.035  & 100 & 0.0020(6)& 0.75(2)&0.05(40)& 0.6938(5) & $2(20)\times 10^{-5}$  & 0.42(2)\\
\hline
$-0.1$ & 0.0525 & 200 & 0.297(9) & 0.71(2)&0.602(2)  & 0.6844(9)& $9(2)\times 10^{-3}$  & 0.19(5)\\
$-0.1$ & 0.05   & 100 & 0.37(1)  & 0.67(1)&0.609(4)  & 0.712(1) & $4.8(7)\times 10^{-3}$& 0.33(5)\\
$-0.1$ & 0.0475 & 100 & 0.094(4) & 0.71(2)&0.562(7)  & 0.7240(8)& $1.7(3)\times 10^{-3}$& 0.36(7)\\
$-0.1$ & 0.045  & 100 & 0.092(5) & 0.69(2)&0.590(9)  & 0.7380(8)& $1.3(2)\times 10^{-3}$& 0.42(5)\\
$-0.1$ & 0.0425 & 100 & 0.031(3) & 0.85(3)&0.4(1)   & 0.7503(8)& $5.6(10)\times 10^{-4}$& 0.46(4)\\
$-0.1$ & 0.04   & 100 & 0.0034(7)& 0.83(2)&0.08(40)&0.7582(7)&$5(30)\times 10^{-5}$      & 0.48(4)\\
\hline
$+0.1$ & 0.0425 & 100 & 0.40(2)  & 0.58(1)&0.493(3) & 0.5620(9) & $5(1)\times 10^{-3}$  & 0.11(3)\\
$+0.1$ & 0.04   & 100 & 0.23(1)  & 0.62(3)&0.487(2) & 0.5877(9) & $1.6(4)\times 10^{-3}$& 0.19(5)\\
$+0.1$ & 0.0375 & 100 & 0.068(5) & 0.70(2)&0.40(4)  & 0.6068(5) & $8(1)\times 10^{-4}$  & 0.24(3)\\
$+0.1$ & 0.035  & 100 & 0.015(2) & 0.72(2)&0.17(26)  & 0.6157(6) & $1.7(40)\times 10^{-4}$& 0.28(2)\\
\hline
\end{tabular}
\end{center}
}
\end{table*}

In order to determine the critical point $(T_c^*,\rho_c^*)$ we empirically 
fit the binodals using the ``diameters'' $(\rho_g^*+\rho_l^*)/2$ equation
\cite{Sengers} 
\bq \nonumber 
\frac{\rho_g^*+\rho_l^*}{2}&=&\rho_c^*+A|T^*-T_c^*|+C|T^*-T_c^*|^{2\beta_I}\\
&&+D|T^*-T_c^*|^{1-\alpha_I}~,
\eq
and the form of the 
Wegner expansion \cite{Wegner1972,Sengers} for the width of the
coexistence curve  
\bq \nonumber
\rho_l^*-\rho_g^*&=&B|T^*-T_c^*|^{\beta_I}+B_1|T^*-T_c^*|^{\beta_I+\Delta_I}\\
&&+B_2|T^*-T_c^*|^{\beta_I+2\Delta_I}~,
\eq
where $A,C,D$, and $B,B_1,B_2$ are coefficients which we take as fitting
parameters as well as $\rho_c^*, T_c^*$. We stress that 
our data do not extend
sufficiently close to the critical region to allow quantitative  
estimates of
critical exponents and non universal quantities, 
still we used the above functional forms as convenient fitting formulae, 
able to capture the typical flatness of the fluid coexistence curves \cite{Frenkel-Smit}.
The pure RPM is believed
\cite{Caillol1996,Luijten2002,Caillol2004,Sengers2009} to
belong to the three-dimensional Ising universality class so we
choose $\beta_I=0.325, \alpha_I=0.11,$ and $\Delta_I=0.51$. 
We are then able to fit the pure RPM case, $\Delta=0$, for which we
find the critical point at $\rho_c^*=0.0319$ and $T_c^*=0.0476$,
the RPM with positive nonadditivity, $\Delta=+0.1$, for which the
critical point is found at $\rho_c^*=0.0275$, $T_c^*=0.0432$, and
the RPM with negative nonadditivity, $\Delta=-0.1$, for which
$\rho_c^*=0.0495$, $T_c^*=0.0526$.  We stress that these numbers, in particular the values of critical densities should be considered more as indicative of the dependence of the critical point location on diamater non additivity than as accurate estimates.

\begin{figure}[htbp]
\begin{center}
\includegraphics[width=10cm]{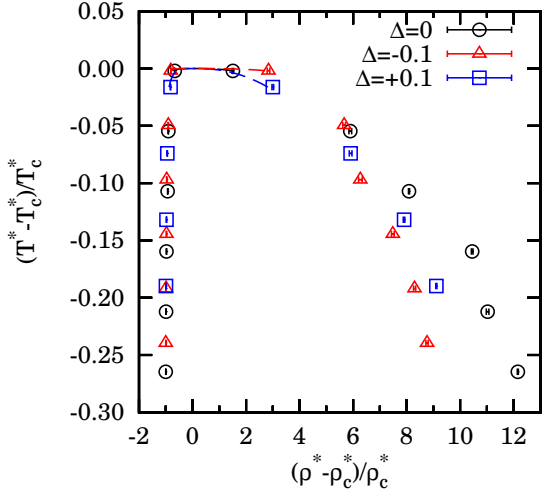}
\end{center}  
\caption{(Color online) Corresponding state representation of the
  phase diagram.} 
\label{fig:cs}
\end{figure}
\begin{figure}[htbp]
\begin{center}
\includegraphics[width=8cm]{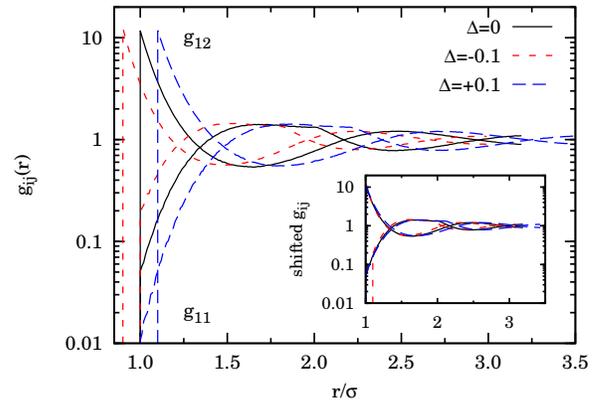}
\end{center}  
\caption{(Color online) RDF on corresponding states of the
  liquid branch at $T^*=0.0375, \rho^*=0.38, \Delta=0$, $T^*=0.035, 
  \rho^*=0.28, \Delta=+0.1$, and $T^*=0.0425, \rho^*=0.46,
  \Delta=-0.1$. In the inset are shown the functions shifted by
  $\pm\Delta$.} 
\label{fig:gr-gemc}
\end{figure}

We believe that our results can be relevant for the
interpretation of experimental work on the phase diagrams of room 
temperature ionic liquids \cite{Saracsan2006} like the phosphonium
halogenide in alkanes solvents and 1-hexyl 3-methyl imidazolium
tetrafluoro borate (C$_6$mimBF$_4$) in alcohols and water. The degree
of nonadditivity seems directly related to the 
anion-cation contact-pairing affinity \cite{Kalcher2010}. The salts
in the (hydrocarbon) solution dissociate in cations (the phosphonium)
and anions (the halogen atoms). The contact affinity between anions
and cations is mediated by the solvent and different solvents produce
different affinities. As a consequence, in the experimental work of
Ref. \cite{Saracsan2006} they observe liquid-liquid coexistence curves
which, depending on the kind of solvent used in the ionic liquid
mixture, can be above (C$_6$mimBF$_4$ in alcohols and water) the one of
the pure RPM theoretical model or below (phosphonium halogenide in
alkanes) in reduced units. Moreover, when plotted into a corresponding state
representation all the experimental binodals seem to collapse on a
same curve even if this occurs very close to the critical point.   
We then try to see if the law of corresponding states holds or not
for our fluid and we find that far from the critical point it is not
strictly satisfied, as shown by Fig. \ref{fig:cs}. Interestingly
enough, a plot of the RDF between corresponding states shows an almost
complete  
overlap of the three curves upon a shift by $\pm\Delta$ in $r$, as is
show by Fig \ref{fig:gr-gemc}. We think that the only visible
difference, the contact values of the like RDF, is a direct hallmark
of the break-up of the corresponding 
states, as physical consequence of the existence of a third relevant
interaction parameter, in addition to the unlike hard-sphere
diameter and the electric charge. 
While the Cl$^-$ ion and the BF$^-_4$ anion may reasonably
well be approximated by a sphere so that the center of charge
is identical with the center of mass, the NTF$^-_2$ anion is by
no means spherical. The NTF$^-_2$ anion is flexible and allows
for different conformers. The nitrogen atom in the
anion is not necessary identical with the center of mass and
the center of charges \cite{Schroer2009}. In these cases instead of
the RPM it is better to choose the PM with ions of
differing sizes as reference system
\cite{Enrique2000,Panagiotopoulos2002,Yan2001}.
%

\section{Theoretical remarks on the clustering}
\label{sec:theory}

Under highly diluted conditions \cite{Valeriani2010} we can
approximate the fluid as an ideal mixture of
cation-anions, anions, and cations with partial densities
$\rho_\pm=(1-\alpha)\rho/2$, $\rho_-=\rho_+=\alpha\rho/2$,
respectively, and for the chemical potentials 
$\mu_\pm=k_BT\ln((1-\alpha)\rho\Lambda_+^3\Lambda_-^3/2K)$,
$\mu_-=k_BT\ln(\alpha\rho\Lambda_+^3/2)$, and
$\mu_+=k_BT\ln(\alpha\rho\Lambda_-^3/2)$, where $\Lambda_-$ and
$\Lambda_+$ are the de Broglie thermal wavelengths of the anions and
cations respectively. $K$ is the configurational integral of a
cation-anion pair
\bq \label{K}
K=4\pi\int_{\sigma(1+\Delta)}^{r_c}r^2e^{\lambda_B/r}\,dr~,
\eq
where $\lambda_B=\sigma/T^*$ is the Bjerrum length and $r_c$ is a
cutoff radius conventionally chosen equal to $\lambda_B/2$
corresponding to the minimum of the integrand. At equilibrium
$\mu_\pm=\mu_++\mu_-$, which implies
$(1-\alpha)/\alpha^2=K\rho/2$. Solving for $\alpha$ yields
\bq
\alpha=\frac{\sqrt{1+2K\rho}-1}{K\rho}~,
\eq
An approximate closed form expression for $K$ valid at low $T^*$ can
be obtained by writing for the anion-cation pair distance
$r=\sigma(1+\Delta)+\delta r$ with $\delta r$ small. Then
$\sigma/r\approx 1/(1+\Delta)-\sigma\delta
r/\sigma^2(1+\Delta)^2=2/(1+\Delta)-r/\sigma(1+\Delta)^2$. Substituting
into Eq. (\ref{K}) and performing the integral with $r_c=\infty$ yields
\bq \nonumber
K&\approx& 4\pi\sigma^3(1+\Delta)^4e^{1/{T^*(1+\Delta)}}T^*\\
&&\times\{1+2(1+\Delta)T^*[1+(1+\Delta)T^*]\}~.
\eq  

\begin{figure}[htbp]
\begin{center}
\includegraphics[width=8cm]{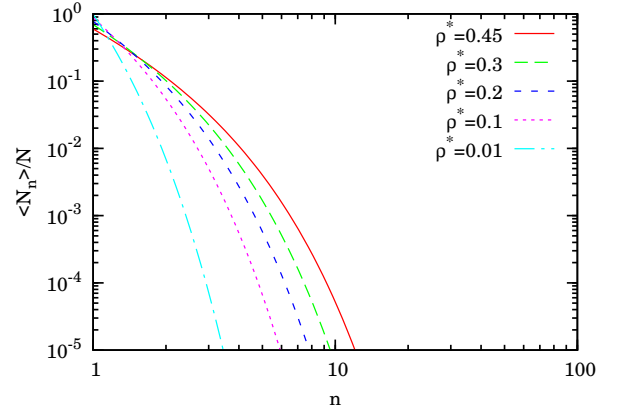}
\end{center}  
\caption{(Color online) Cluster analysis described in the text for
  $n^c=100$ at various densities.} 
\label{fig:tani-1}
\end{figure}

In our simulations we are never in this very diluted condition and as
a consequence we observe the formation of clusters of an higher number
of particles than just the dimers. So to estimate the
cluster concentrations $x^c_n=\langle N_n\rangle/N$, we need a
different analysis closer in spirit to the one of Tani and Henderson
\cite{Tani1983,Fantoni2011,*Fantoni2012}. Simplifying that analysis we
can consider as the 
inter-cluster configurational partition function the one of an ideal
gas of clusters, in reduced units, $Z_{inter}\approx (V/\sigma^3)^{N_t}$,
where $N_t=\sum_{n=1}^{n^c} N_n$ is the total number of clusters and
we assume to have only clusters made of up to $n^c$ particles. Then
the equations for the equilibrium cluster concentrations $x^c_n$ are
\bq \label{xcn}
x^c_n&=&\lambda^nz^{intra}_n/\rho^*~,~~~n=1,2,\ldots,n^c~,\\
1&=&\sum_{n=1}^{n^c} nx^c_n~,
\eq
where $z_n^{intra}$ are the configurational intra-cluster partition
functions in reduced units with $z_1^{intra}=2$ and
$\lambda(=\alpha\rho^*/2)$ is a Lagrange multiplier.  
Moreover, neglecting the excess internal energy of the clusters we can
approximate $z^{intra}_{n}\approx
(v_n/\sigma^3)^{n-1}\sum_{s=0}^n(s!(n-s)!)^{-1}=(v_n/\sigma^3)^{n-1}2^n/n!$
where $v_n$ is the  
volume of an $n-$cluster. Moreover, assuming further the cluster to be in a closed
packed configuration we can approximate, for $\Delta=0$, $v_n\approx
n\sigma^3/\sqrt{2}$. Notice that for $\Delta\neq 0$ we would expect
$v_n$ to change by a constant multiplicative factor which would still
give the same result for the cluster concentrations. Clearly a proper
analysis of the $n-$cluster volume would require a MC simulation
\cite{Gillan1983}. This temperature independent approximation gives
for $n^c=100$ the results shown in Fig. \ref{fig:tani-1} (note that the
results have very small dependence on $n^c$).

Form the figure we can say that our simulation results for
$T^*=0.1$ and $\Delta=-0.3$ have qualitatively the same
behavior of our oversimplified analysis. This justifies the fit of
Table \ref{tab:ab} where the Laplace multiplier is considered as a
fitting parameter. The strong
dependence from the nonadditivity (and on temperature) that we observe
in the simulation is 
an indication that the approximation of neglecting the excess internal
energy of a cluster is too severe. One should consider
$z^{intra}_n=e^{-nf_n^\text{ex}/T^*}(v_n/\sigma^3)^{n-1}2^n/n!$, where 
$f_n^\text{ex}(T^*)=\int_0^{1/T^*}u_n^\text{ex}(1/x)\,dx$ is the 
excess free energy per particle of the $n-$cluster and
$u_n^\text{ex}(T^*)=(\epsilon\sigma/q^2)\langle\sum_{i>j=1}^n
\phi_{ij}(r_{ij})\rangle/n$  
is the reduced excess internal energy per particle of the
$n-$cluster. Note once 
again that choosing an $f_n^\text{ex}$ independent of $n$ would lead to the
same oversimplified result we described for the cluster
concentrations. What really matters is the combined dependence of
$f_n^\text{ex}(T^*)$ on $n$ and $T^*$ which can be assessed within the
MC simulation \cite{Gillan1983,Fantoni2011,*Fantoni2012}. For example
the curves of Figs. \ref{fig:ncl-t0.1} and \ref{fig:ncl-t0.04} with
percolating clusters are better fitted by the following 
three parameters expression $x_n^c\approx \lambda^{n+an^2}n^{bn}/n!$.

One thing that can be done is to
distinguish amongst the clusters of $n$ particles between the ones
formed by $s$ negative particles and $t$ positive particles with
$t+s=n$, as done in Ref. \cite{Caillol1995}, in order to be able to
approximate analytically the intra-cluster excess free energy per
particle
\bq 
z_{n}^{intra}&=&\sum_{s=0}^n z_{s,n-s}^{intra}~,\\ \nonumber
z_{s,t}^{intra}&=&\frac{1}{s!t!}\frac{1}{\sigma^{3(s+t-1)}}\int_{\Omega_{s,t}}
d\rr_2\ldots d\rr_{s+t}\\ \label{zintra}
&&\times e^{-\beta\sum_{\mu>\nu=1}^{s+t}\phi_{i_\mu j_\nu}(r_{\mu\nu})}~,
\eq
where the configurational integral goes only over the relative
positions and it covers the region $\Omega_{s,t}$ of cluster configuration
space. 
This way one can quantitatively \cite{Caillol1995} estimate
how the Tani and Henderson theory \cite{Tani1983} deviates from the
exact MC results. 

We immediately see how $z_{1,1}^{intra}\propto K/\sigma^3$ becomes
bigger and bigger as $\Delta\to -1$ and the same holds for all the
$z_{k,k}^{intra}$ which clearly dominate over all the others
$z_{s,t}^{intra}$ with $s\neq t$. This qualitatively explains the
Fig. \ref{fig:dip} as is shown in Fig. \ref{fig:tani-2} where we show
the results from the approximation described in Appendix \ref{app:1}
for $n^c=30, T^*=0.1, \rho^*=0.45,a=1.5,b=0.9$ and various values of
$\Delta$ (note that the results have very small dependence on $n^c$). 

\begin{figure}[htbp]
\begin{center}
\includegraphics[width=8cm]{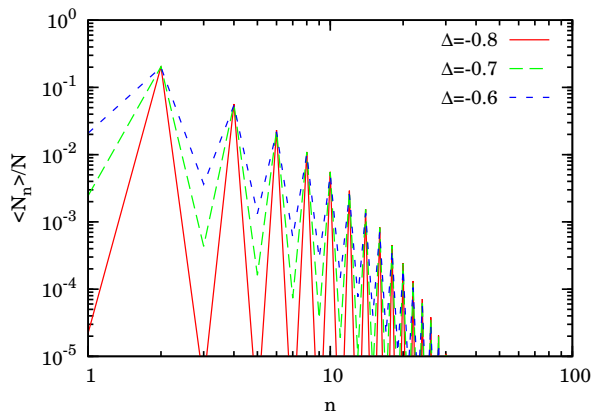}
\end{center}  
\caption{(Color online) Cluster analysis described in the text for
  $n^c=30, T^*=0.1, \rho^*=0.45,a=1.5,b=0.9$ at various values of
  $\Delta$.}  
\label{fig:tani-2}
\end{figure}
%

\section{Conclusions}
\label{sec:conclusions}

In conclusion we have performed $NVT$ MC simulations of the RPM with
non-additive hard-sphere diameters outside the coexistence region with
particular emphasis on the clustering properties. In order to
establish whether the cluster analysis falls outside the gas-liquid
coexistence region for a given value of the nonadditivity we
accurately determined the binodals of the non-additive fluid using the
Gibbs ensemble method after a density distribution function analysis
to get a first
insights on the shifts of the coexistence region with the
nonadditivity. It turned out that a negative nonadditivity tends to
shift to higher temperatures and higher densities the critical point
whether a positive one shifts it to lower temperatures and
densities. The law of corresponding states does not seem to be
strictly fulfilled over an extended region below the critical
point for $\Delta=0,\pm 
0.1$. Our results can be used as a theoretical support to the
analysis of experimental work on room temperature ionic liquids
\cite{Kleemeier1999,Saracsan2006,Schroer2009} where shifts in the
liquid-liquid binodals akin to ours are observed as a function of the
kind of solvent used in the ionic mixture.

From the cluster analysis, we were able to distinguish between two
kind of behaviors for the cluster concentrations. When we do not
observe percolating clusters during the simulation, the curves for the
cluster concentrations as a function of the cluster size are
independent of the number of particles used in the simulation. When we
observe percolation during the simulation the 
curves depend on the number of particles used in the simulation \gio{but}
obey a straightforward scaling with $N$ relationship. 

At low densities the negative non-additive fluid has stronger
clustering than the pure RPM whereas at high densities is the positive
non-additive fluid to have the strongest one. The positive
non-additive fluid is the first one reaching the percolating
clusters upon an increase of density. This certainly depends on
the fact that for a positive nonadditivity the ions have less space
where to move at a given density and, due to he presence of 
two opposite charged species, 
there is a competition between the tendency to clustering
driven by the Coulomb interaction and tendency to demixing
due to entropic reasons. A negative nonadditivity tends to favor the
formation of the neutrally charged clusters starting with dipolar
ones. Traces of these features can also be read from an analysis of 
the partial radial distribution function and structure factors. Our
clustering results can be summarized observing that  
at high density for a positive deviations from additivity we have more
clustering than in the additive model whereas for a negative deviation
from additivity we have less clustering than in the additive model. 
At
low density the reverse behavior is found. These results can be
explained by the following arguments: at high density a positive
nonadditivity leaves less effective volume to the 
particles and a negative nonadditivity leaves more effective volume
relative to the additive model; at low density a
negative nonadditivity favors the formation of neutral clusters and a
positive nonadditivity favors the competition between the
tendency to demixing in a neutral mixture and the tendency to
microscopic intermixing of the two species favored by the Coulombic
interactions. These observations are in agreement with the fact that
the energy of a cation-anion pair at contact increases for positive
nonadditivity and decreases for negative nonadditivity. 

A simple, temperature independent, clustering theory where we regard 
the clusters as forming an ideal gas and we approximate the
$n-$cluster as an ideal ensemble of $n$ particles in a closed packed
configuration can be used to qualitatively explain the cluster
concentrations observed at not to high density and absolute value of
the nonadditivity. In order to qualitatively explain the prevalence of
the neutral clusters in the negatively non-additive fluid it is
necessary to refine the approximation at the intracluster level.

In the future it would be desirable to make quantitative the
comparison between clustering theory and MC exact results. 
It would be also interesting the determination of the percolation
threshold as a function of nonadditivity. 
In the temperature density phase diagram, one can determine the
percolation threshold calculating the fraction of configurations with
percolating clusters within the $NVT$ simulation as a function of
density for two systems of different size $N$. A point of the
percolation threshold results then from where the curves of
the two systems meet.    

\appendix
\section{Approximated intracluster configurational partition function
  for negative nonadditivity} 
\label{app:1}

Let us call the anions $i_-=1_-,\ldots,s_-$ and the cations
$j_+=1_+,\ldots,t_+$. From Eq. (\ref{zintra}) in the main text follows 
\bq \nonumber
z_{t,t}^{intra}&=&\frac{1}{t!^2}\frac{1}{\sigma^{3(2t-1)}}\int_{\Omega_{t,t}}
\prod_{l=2}^td\rr_{1_+l_-}\prod_{k=1}^td\rr_{k_+k_-}\\ \nonumber
&&\times \prod_{i>j=1}^te^{-2\lambda_B/r_{i_+j_+}}
\prod_{i,j=1}^te^{+\lambda_B/r_{i_+j_-}}\\
&\approx&\frac{1}{t!^2}\frac{1}{\sigma^{3(2t-1)}}\int_{\Omega_{t,t}}
\prod_{l=2}^td\rr_{1_+l_-}\prod_{k=1}^td\rr_{k_+k_-}\\ \nonumber
&&\times \prod_{i,j=1}^te^{+\lambda_B/r_{i_+j_-}}~,
\eq
where we approximated $e^{-\lambda_B/r}\approx 1$ which is justified
at high $T^*<1/2(1+\Delta)$ or low $\lambda_B$. Now we observe that
for example $r_{1_+2_-}=|\rr_{1_+1_-}+\rr_{1_-2_-}|$ with
$r_{1_-2_-}>\sigma$ and $e^{+\lambda_B/r_{1_+2_-}}\approx 1$. So that for
negative nonadditivity we can further approximate   
\bq \nonumber
z_{t,t}^{intra}&\approx&\frac{1}{t!^2}\frac{1}{\sigma^{3(2t-1)}}\int_{\Omega_{t,t}}
\prod_{l=2}^td\rr_{1_+l_-}\prod_{k=1}^td\rr_{k_+k_-}\\ \nonumber
&&\times\prod_{i,j=1}^te^{+\lambda_B/r_{i_+j_-}}\\ \nonumber
&\approx&\frac{1}{t!^2}\frac{1}{\sigma^{3(2t-1)}}\int_{\Omega_{t,t}}
\prod_{l=2}^td\rr_{1_+l_-}\prod_{k=1}^td\rr_{k_+k_-}\\ \nonumber
&&\times\prod_{i=1}^te^{+\lambda_B/r_{i_+i_-}}\\
&\stackrel{\mbox{\normalsize  $\propto$}}{\sim}&
\frac{(2t)^{b(2t-1)}}{t!^2}(K/K_0)^t~,
\eq
where the factor $(2t)^{b(2t-1)}$ takes into account the volume of
$\Omega_{t,t}$ with $b$ a free parameter and
\bq
K/K_0=\int_{a\sigma(1+\Delta)}^{\lambda_B/2}r^2e^{+\lambda_B/r}\,dr/
\int_{a\sigma(1+\Delta)}^{\lambda_B/2}r^2\,dr~,
\eq
with $a$ a second free parameter. With the same approximations we can say
\bq
z_{s,t}^{intra}\stackrel{\mbox{\normalsize $\propto$}}{\sim}
\frac{(s+t)^{b(s+t-1)}}{s!t!}(K/K_0)^{\min\{s,t\}}~. 
\eq

\begin{acknowledgments}
R.F. would like to acknowledge the use of the computational facilities
of CINECA through the ISCRA call. The GEMC code took $\sim 26$ min of
CPU time for 10 million steps of a system of size $N=200$ on the IBM
PLX (iDataPlex DX360M3) cluster. 
\end{acknowledgments}

\begin{thebibliography}{81}
\expandafter\ifx\csname natexlab\endcsname\relax\def\natexlab#1{#1}\fi
\expandafter\ifx\csname bibnamefont\endcsname\relax
  \def\bibnamefont#1{#1}\fi
\expandafter\ifx\csname bibfnamefont\endcsname\relax
  \def\bibfnamefont#1{#1}\fi
\expandafter\ifx\csname citenamefont\endcsname\relax
  \def\citenamefont#1{#1}\fi
\expandafter\ifx\csname url\endcsname\relax
  \def\url#1{\texttt{#1}}\fi
\expandafter\ifx\csname urlprefix\endcsname\relax\def\urlprefix{URL }\fi
\providecommand{\bibinfo}[2]{#2}
\providecommand{\eprint}[2][]{\url{#2}}

\bibitem[{\citenamefont{{D. Henderson, M. Holovko, and A.
  Trokhymchuk}}(2004)}]{Henderson2004}
\bibinfo{editor}{\bibnamefont{{D. Henderson, M. Holovko, and A. Trokhymchuk}}},
  ed., \emph{\bibinfo{title}{Ionic Soft Matter: Modern Trends in Theory and
  Applications}}, vol. \bibinfo{volume}{206} of \emph{\bibinfo{series}{NATO
  Science series}} (\bibinfo{publisher}{Springer},
  \bibinfo{address}{Netherlands}, \bibinfo{year}{2004}).

\bibitem[{\citenamefont{{P. G. de Gennes}}(1992)}]{deGennes1992}
\bibinfo{author}{\bibnamefont{{P. G. de Gennes}}}, \bibinfo{journal}{Rev. Mod.
  Phys.} \textbf{\bibinfo{volume}{64}}, \bibinfo{pages}{645}
  (\bibinfo{year}{1992}).

\bibitem[{\citenamefont{{R. Fantoni and G. Pastore}}(2004)}]{Fantoni04b}
\bibinfo{author}{\bibnamefont{{R. Fantoni and G. Pastore}}},
  \bibinfo{journal}{J. Chem. Phys.} \textbf{\bibinfo{volume}{120}},
  \bibinfo{pages}{10681} (\bibinfo{year}{2004}).

\bibitem[{\citenamefont{{D. Gazzillo, A. Giacometti, R. Fantoni, and P.
  Sollich}}(2006)}]{Fantoni06a}
\bibinfo{author}{\bibnamefont{{D. Gazzillo, A. Giacometti, R. Fantoni, and P.
  Sollich}}}, \bibinfo{journal}{Phys. Rev. E} \textbf{\bibinfo{volume}{74}},
  \bibinfo{pages}{051407} (\bibinfo{year}{2006}).

\bibitem[{\citenamefont{{R. Fantoni, A. Giacometti, A. Malijevsk\'y, and A.
  Santos}}(2010)}]{Fantoni10a}
\bibinfo{author}{\bibnamefont{{R. Fantoni, A. Giacometti, A. Malijevsk\'y, and
  A. Santos}}}, \bibinfo{journal}{J. Chem. Phys.}
  \textbf{\bibinfo{volume}{133}}, \bibinfo{pages}{024101}
  (\bibinfo{year}{2010}).

\bibitem[{\citenamefont{{R. Fantoni, A. Giacometti, F. Sciortino, and G.
  Pastore}}(2011)}]{Fantoni11a}
\bibinfo{author}{\bibnamefont{{R. Fantoni, A. Giacometti, F. Sciortino, and G.
  Pastore}}}, \bibinfo{journal}{Soft Matter} \textbf{\bibinfo{volume}{7}},
  \bibinfo{pages}{2419} (\bibinfo{year}{2011}).

\bibitem[{\citenamefont{{R. Fantoni}}(2012{\natexlab{a}})}]{Fantoni12a}
\bibinfo{author}{\bibnamefont{{R. Fantoni}}}, \bibinfo{journal}{Eur. Phys. J.
  B} \textbf{\bibinfo{volume}{85}}, \bibinfo{pages}{108}
  (\bibinfo{year}{2012}{\natexlab{a}}).

\bibitem[{\citenamefont{{R. Fantoni, J. W. O. Salari, and B.
  Klumperman}}(2012)}]{Fantoni12c}
\bibinfo{author}{\bibnamefont{{R. Fantoni, J. W. O. Salari, and B.
  Klumperman}}}, \bibinfo{journal}{Phys. Rev. E} \textbf{\bibinfo{volume}{85}},
  \bibinfo{pages}{061404} (\bibinfo{year}{2012}).

\bibitem[{\citenamefont{{J. P. Hansen and I. R. McDonald}}(1986)}]{Hansen}
\bibinfo{author}{\bibnamefont{{J. P. Hansen and I. R. McDonald}}},
  \emph{\bibinfo{title}{Theory of Simple Liquids}}
  (\bibinfo{publisher}{Academic Press}, \bibinfo{year}{1986}),
  \bibinfo{edition}{2nd} ed.

\bibitem[{\citenamefont{{G. Stell, K. C. Wu, and B. Larsen}}(1976)}]{Stell1976}
\bibinfo{author}{\bibnamefont{{G. Stell, K. C. Wu, and B. Larsen}}},
  \bibinfo{journal}{Phys. Rev. Lett.} \textbf{\bibinfo{volume}{37}},
  \bibinfo{pages}{1369} (\bibinfo{year}{1976}).

\bibitem[{\citenamefont{{T. L. Croxton and D. A.
  McQuarries}}(1979)}]{Croxton1979}
\bibinfo{author}{\bibnamefont{{T. L. Croxton and D. A. McQuarries}}},
  \bibinfo{journal}{J. Phys. Chem.} \textbf{\bibinfo{volume}{83}},
  \bibinfo{pages}{1840} (\bibinfo{year}{1979}).

\bibitem[{\citenamefont{{M. J. Gillan}}(1983)}]{Gillan1983}
\bibinfo{author}{\bibnamefont{{M. J. Gillan}}}, \bibinfo{journal}{Mol. Phys.}
  \textbf{\bibinfo{volume}{49}}, \bibinfo{pages}{421} (\bibinfo{year}{1983}).

\bibitem[{\citenamefont{{T. Cartailler, P. Turq, L. Blum, and N.
  Condamine}}(1992)}]{Cartailler1992}
\bibinfo{author}{\bibnamefont{{T. Cartailler, P. Turq, L. Blum, and N.
  Condamine}}}, \bibinfo{journal}{J. Phys. Chem.}
  \textbf{\bibinfo{volume}{96}}, \bibinfo{pages}{6766} (\bibinfo{year}{1992}).

\bibitem[{\citenamefont{{J. A. Given}}(1992)}]{Given1992}
\bibinfo{author}{\bibnamefont{{J. A. Given}}}, \bibinfo{journal}{Phys. Rev. A}
  \textbf{\bibinfo{volume}{45}}, \bibinfo{pages}{3849} (\bibinfo{year}{1992}).

\bibitem[{\citenamefont{{J. A. Given and G. Stell}}(1992)}]{Given1992b}
\bibinfo{author}{\bibnamefont{{J. A. Given and G. Stell}}},
  \bibinfo{journal}{J. Chem. Phys.} \textbf{\bibinfo{volume}{96}},
  \bibinfo{pages}{9233} (\bibinfo{year}{1992}).

\bibitem[{\citenamefont{{M. E. Fisher and Y. Levin}}(1993)}]{Fisher1993}
\bibinfo{author}{\bibnamefont{{M. E. Fisher and Y. Levin}}},
  \bibinfo{journal}{Phys. Rev. Lett.} \textbf{\bibinfo{volume}{71}},
  \bibinfo{pages}{3826} (\bibinfo{year}{1993}).

\bibitem[{\citenamefont{{M. E. Fisher}}(1994)}]{Fisher1994}
\bibinfo{author}{\bibnamefont{{M. E. Fisher}}}, \bibinfo{journal}{J. Stat.
  Phys.} \textbf{\bibinfo{volume}{75}}, \bibinfo{pages}{1}
  (\bibinfo{year}{1994}).

\bibitem[{\citenamefont{{G. Stell}}(1995)}]{Stell1995}
\bibinfo{author}{\bibnamefont{{G. Stell}}}, \bibinfo{journal}{J. Stat. Phys.}
  \textbf{\bibinfo{volume}{78}}, \bibinfo{pages}{197} (\bibinfo{year}{1995}).

\bibitem[{\citenamefont{{Y. Zhou, S. Yeh, and G. Stell}}(1995)}]{Zhou1995}
\bibinfo{author}{\bibnamefont{{Y. Zhou, S. Yeh, and G. Stell}}},
  \bibinfo{journal}{J. Chem. Phys.} \textbf{\bibinfo{volume}{102}},
  \bibinfo{pages}{5785} (\bibinfo{year}{1995}).

\bibitem[{\citenamefont{{S. G. Yeh, Y. Q. Zhou, and G. Stell}}(1996)}]{Yeh1996}
\bibinfo{author}{\bibnamefont{{S. G. Yeh, Y. Q. Zhou, and G. Stell}}},
  \bibinfo{journal}{J. Phys. Chem.} \textbf{\bibinfo{volume}{100}},
  \bibinfo{pages}{1415} (\bibinfo{year}{1996}).

\bibitem[{\citenamefont{{J. A. Given and G. Stell}}(1997)}]{Given1997}
\bibinfo{author}{\bibnamefont{{J. A. Given and G. Stell}}},
  \bibinfo{journal}{J. Chem. Phys.} \textbf{\bibinfo{volume}{106}},
  \bibinfo{pages}{1195} (\bibinfo{year}{1997}).

\bibitem[{\citenamefont{{J. Jiang, L. Blum, O. Bernard, J. M. Prausnitz, and S.
  I. Sandler}}(2002)}]{Jiang2002}
\bibinfo{author}{\bibnamefont{{J. Jiang, L. Blum, O. Bernard, J. M. Prausnitz,
  and S. I. Sandler}}}, \bibinfo{journal}{J. Chem. Phys.}
  \textbf{\bibinfo{volume}{116}}, \bibinfo{pages}{7977} (\bibinfo{year}{2002}).

\bibitem[{\citenamefont{{H. L. Friedman and B. Larsen}}(1979)}]{Larsen1979}
\bibinfo{author}{\bibnamefont{{H. L. Friedman and B. Larsen}}},
  \bibinfo{journal}{J. Chem. Phys.} \textbf{\bibinfo{volume}{70}},
  \bibinfo{pages}{92} (\bibinfo{year}{1979}).

\bibitem[{\citenamefont{{P. N. Vorontsov-Veliaminov, A. M. El\`yashevich, L. A.
  Morgenshtern, and V. P. Chasovshikh}}(1976)}]{Veliaminov1976}
\bibinfo{author}{\bibnamefont{{P. N. Vorontsov-Veliaminov, A. M. El\`yashevich,
  L. A. Morgenshtern, and V. P. Chasovshikh}}}, \bibinfo{journal}{High. Temp.
  (USSR)} \textbf{\bibinfo{volume}{8}}, \bibinfo{pages}{261}
  (\bibinfo{year}{1976}).

\bibitem[{\citenamefont{{V. P. Chasovshikh and P. N.
  Vorontsov-Veliaminov}}(1976)}]{Chasovshikh1976}
\bibinfo{author}{\bibnamefont{{V. P. Chasovshikh and P. N.
  Vorontsov-Veliaminov}}}, \bibinfo{journal}{High. Temp. (USSR)}
  \textbf{\bibinfo{volume}{14}}, \bibinfo{pages}{174} (\bibinfo{year}{1976}).

\bibitem[{\citenamefont{{A. Z. Panagiotopoulos}}(1992)}]{Panagiotopoulos92}
\bibinfo{author}{\bibnamefont{{A. Z. Panagiotopoulos}}},
  \bibinfo{journal}{Fluid. Phase Equil.} \textbf{\bibinfo{volume}{76}},
  \bibinfo{pages}{97} (\bibinfo{year}{1992}).

\bibitem[{\citenamefont{{I. S. Graham and J. P. Valleau}}(1990)}]{Graham1990}
\bibinfo{author}{\bibnamefont{{I. S. Graham and J. P. Valleau}}},
  \bibinfo{journal}{J. Phys. Chem.} \textbf{\bibinfo{volume}{94}},
  \bibinfo{pages}{7894} (\bibinfo{year}{1990}).

\bibitem[{\citenamefont{{J.-M. Caillol}}(1994)}]{Caillol1994}
\bibinfo{author}{\bibnamefont{{J.-M. Caillol}}}, \bibinfo{journal}{J. Chem.
  Phys.} \textbf{\bibinfo{volume}{100}}, \bibinfo{pages}{2161}
  (\bibinfo{year}{1994}).

\bibitem[{\citenamefont{{G. Orkoulas and A. Z.
  Panagiotopoulos}}(1994)}]{Orkoulas1994}
\bibinfo{author}{\bibnamefont{{G. Orkoulas and A. Z. Panagiotopoulos}}},
  \bibinfo{journal}{J. Chem. Phys.} \textbf{\bibinfo{volume}{101}},
  \bibinfo{pages}{1452} (\bibinfo{year}{1994}).

\bibitem[{\citenamefont{{J.-M. Caillol and J.-J. Weis}}(1995)}]{Caillol1995}
\bibinfo{author}{\bibnamefont{{J.-M. Caillol and J.-J. Weis}}},
  \bibinfo{journal}{J. Chem. Phys.} \textbf{\bibinfo{volume}{102}},
  \bibinfo{pages}{7610} (\bibinfo{year}{1995}).

\bibitem[{\citenamefont{{G. Orkoulas and A. Z.
  Panagiotopoulos}}(1999)}]{Orkoulas1999}
\bibinfo{author}{\bibnamefont{{G. Orkoulas and A. Z. Panagiotopoulos}}},
  \bibinfo{journal}{J. Chem. Phys.} \textbf{\bibinfo{volume}{110}},
  \bibinfo{pages}{1581} (\bibinfo{year}{1999}).

\bibitem[{\citenamefont{{Q. Yan and J. J. de Pablo}}(1999)}]{Yan1999}
\bibinfo{author}{\bibnamefont{{Q. Yan and J. J. de Pablo}}},
  \bibinfo{journal}{J. Chem. Phys.} \textbf{\bibinfo{volume}{111}},
  \bibinfo{pages}{9509} (\bibinfo{year}{1999}).

\bibitem[{\citenamefont{{E. Luijten, M. E. Fisher, and A. Z.
  Panagiotopoulos}}(2002)}]{Luijten2002}
\bibinfo{author}{\bibnamefont{{E. Luijten, M. E. Fisher, and A. Z.
  Panagiotopoulos}}}, \bibinfo{journal}{Phys. Rev. Lett.}
  \textbf{\bibinfo{volume}{88}}, \bibinfo{pages}{185701}
  (\bibinfo{year}{2002}).

\bibitem[{\citenamefont{{J.-M. Caillol, D. Levesque, and J.-J.
  Weis}}(2002)}]{Caillol2002}
\bibinfo{author}{\bibnamefont{{J.-M. Caillol, D. Levesque, and J.-J. Weis}}},
  \bibinfo{journal}{J. Chem. Phys.} \textbf{\bibinfo{volume}{116}},
  \bibinfo{pages}{10794} (\bibinfo{year}{2002}).

\bibitem[{\citenamefont{{P. J. Camp and G. N. Patey}}(1999)}]{Camp1999}
\bibinfo{author}{\bibnamefont{{P. J. Camp and G. N. Patey}}},
  \bibinfo{journal}{J. Chem. Phys.} \textbf{\bibinfo{volume}{111}},
  \bibinfo{pages}{9000} (\bibinfo{year}{1999}).

\bibitem[{\citenamefont{{J. M. Romero-Enrique, G. Orkoulas, A. Z.
  Panagiotopoulos, and M. E. Fisher}}(2000)}]{Enrique2000}
\bibinfo{author}{\bibnamefont{{J. M. Romero-Enrique, G. Orkoulas, A. Z.
  Panagiotopoulos, and M. E. Fisher}}}, \bibinfo{journal}{Phys. Rev. Lett.}
  \textbf{\bibinfo{volume}{85}}, \bibinfo{pages}{4558} (\bibinfo{year}{2000}).

\bibitem[{\citenamefont{{A. Z. Panagiotopoulos and M. E.
  Fisher}}(2002)}]{Panagiotopoulos2002}
\bibinfo{author}{\bibnamefont{{A. Z. Panagiotopoulos and M. E. Fisher}}},
  \bibinfo{journal}{Phys. Rev. Lett.} \textbf{\bibinfo{volume}{88}},
  \bibinfo{pages}{045701} (\bibinfo{year}{2002}).

\bibitem[{\citenamefont{{Q. Yan and J. J. de Pablo}}(2001)}]{Yan2001}
\bibinfo{author}{\bibnamefont{{Q. Yan and J. J. de Pablo}}},
  \bibinfo{journal}{Phys. Rev. Lett.} \textbf{\bibinfo{volume}{86}},
  \bibinfo{pages}{2054} (\bibinfo{year}{2001}).

\bibitem[{\citenamefont{{Q. Yan and J. J. de Pablo}}(2002)}]{Yan2002}
\bibinfo{author}{\bibnamefont{{Q. Yan and J. J. de Pablo}}},
  \bibinfo{journal}{Phys. Rev. Lett.} \textbf{\bibinfo{volume}{88}},
  \bibinfo{pages}{095504} (\bibinfo{year}{2002}).

\bibitem[{\citenamefont{{G. Pastore, P. V. Giaquinta, J. S. Thakur, and M. P.
  Tosi}}(1986)}]{Pastore1985}
\bibinfo{author}{\bibnamefont{{G. Pastore, P. V. Giaquinta, J. S. Thakur, and
  M. P. Tosi}}}, \bibinfo{journal}{J. Chem. Phys.}
  \textbf{\bibinfo{volume}{84}}, \bibinfo{pages}{1827} (\bibinfo{year}{1986}),
  \bibinfo{note}{{The relationship between our reduced units and theirs is as
  follows: $\rho^*=3\sigma^3/4\pi$ and $T^*=\sigma/\Gamma$.}}

\bibitem[{\citenamefont{{I. Kalcher, J. C. F. Schulz, and J.
  Dzubiella}}(2010)}]{Kalcher2010}
\bibinfo{author}{\bibnamefont{{I. Kalcher, J. C. F. Schulz, and J.
  Dzubiella}}}, \bibinfo{journal}{Phys. Rev. Lett.}
  \textbf{\bibinfo{volume}{104}}, \bibinfo{pages}{097802}
  (\bibinfo{year}{2010}).

\bibitem[{\citenamefont{{M. Rovere and G. Pastore}}(1994)}]{Rovere1994}
\bibinfo{author}{\bibnamefont{{M. Rovere and G. Pastore}}},
  \bibinfo{journal}{J. Phys.: Condens. Matter} \textbf{\bibinfo{volume}{6}},
  \bibinfo{pages}{A163} (\bibinfo{year}{1994}).

\bibitem[{\citenamefont{{E. Lomba, M. Alvarez, L. L. Lee, and N. G.
  Almarza}}(1996)}]{Lomba1996}
\bibinfo{author}{\bibnamefont{{E. Lomba, M. Alvarez, L. L. Lee, and N. G.
  Almarza}}}, \bibinfo{journal}{J. Chem. Phys.} \textbf{\bibinfo{volume}{104}},
  \bibinfo{pages}{4180} (\bibinfo{year}{1996}).

\bibitem[{\citenamefont{{K. Jagannathan and A.
  Yethiraj}}(2003)}]{Jagannathan2003}
\bibinfo{author}{\bibnamefont{{K. Jagannathan and A. Yethiraj}}},
  \bibinfo{journal}{J. Chem. Phys.} \textbf{\bibinfo{volume}{118}},
  \bibinfo{pages}{7907} (\bibinfo{year}{2003}).

\bibitem[{\citenamefont{{W. T. G\'o\'zd\'z}}(2003)}]{Gozdz2003}
\bibinfo{author}{\bibnamefont{{W. T. G\'o\'zd\'z}}}, \bibinfo{journal}{J. Chem.
  Phys.} \textbf{\bibinfo{volume}{119}}, \bibinfo{pages}{3309}
  (\bibinfo{year}{2003}).

\bibitem[{\citenamefont{{A. Buhot}}(2005)}]{Buhot2005}
\bibinfo{author}{\bibnamefont{{A. Buhot}}}, \bibinfo{journal}{J. Chem. Phys.}
  \textbf{\bibinfo{volume}{122}}, \bibinfo{pages}{024105}
  (\bibinfo{year}{2005}).

\bibitem[{\citenamefont{{A. Santos, M. L\'opez de Haro, and S. B.
  Yuste}}(2010)}]{Santos2010}
\bibinfo{author}{\bibnamefont{{A. Santos, M. L\'opez de Haro, and S. B.
  Yuste}}}, \bibinfo{journal}{J. Chem. Phys.} \textbf{\bibinfo{volume}{132}},
  \bibinfo{pages}{204506} (\bibinfo{year}{2010}).

\bibitem[{\citenamefont{{R. Fantoni and G. Pastore}}(2013)}]{Fantoni2013}
\bibinfo{author}{\bibnamefont{{R. Fantoni and G. Pastore}}},
  \bibinfo{journal}{Europhys. Lett.}  (\bibinfo{year}{2013}), \bibinfo{note}{in
  press}.

\bibitem[{\citenamefont{{W. C. K. Poon, S. U. Egelhaaf, J. Stellbrink, J.
  Allgaier, A. B. Schofield, and P. N. Pusey}}(2001)}]{Poon2001}
\bibinfo{author}{\bibnamefont{{W. C. K. Poon, S. U. Egelhaaf, J. Stellbrink, J.
  Allgaier, A. B. Schofield, and P. N. Pusey}}}, \bibinfo{journal}{Phil. Trans.
  R. Soc. Lond. A} \textbf{\bibinfo{volume}{359}}, \bibinfo{pages}{897}
  (\bibinfo{year}{2001}).

\bibitem[{\citenamefont{{W. C. K. Poon}}(2002)}]{Poon2002}
\bibinfo{author}{\bibnamefont{{W. C. K. Poon}}}, \bibinfo{journal}{J. Phys.:
  Condens. Matter} \textbf{\bibinfo{volume}{14}}, \bibinfo{pages}{R859}
  (\bibinfo{year}{2002}).

\bibitem[{\citenamefont{{H. Weing\"artner, M. Kleemeier, S. Wiegand, and W.
  Sch\"oer}}(1995)}]{Weingartner1995}
\bibinfo{author}{\bibnamefont{{H. Weing\"artner, M. Kleemeier, S. Wiegand, and
  W. Sch\"oer}}}, \bibinfo{journal}{J. Stat. Phys.}
  \textbf{\bibinfo{volume}{78}}, \bibinfo{pages}{169} (\bibinfo{year}{1995}).

\bibitem[{\citenamefont{{M. Kleemeier, S. Wiegand, W. Schr\"oer, and H.
  Weing\"artner}}(1999)}]{Kleemeier1999}
\bibinfo{author}{\bibnamefont{{M. Kleemeier, S. Wiegand, W. Schr\"oer, and H.
  Weing\"artner}}}, \bibinfo{journal}{J. Chem. Phys.}
  \textbf{\bibinfo{volume}{110}}, \bibinfo{pages}{3085} (\bibinfo{year}{1999}).

\bibitem[{\citenamefont{{D. Saracsan, C. Rybarsch, and W.
  Schr\"oer}}(2006)}]{Saracsan2006}
\bibinfo{author}{\bibnamefont{{D. Saracsan, C. Rybarsch, and W. Schr\"oer}}},
  \bibinfo{journal}{Z. Phys. Chem.} \textbf{\bibinfo{volume}{220}},
  \bibinfo{pages}{1417} (\bibinfo{year}{2006}).

\bibitem[{\citenamefont{{W. Schr\"oer and V. R. Vale}}(2009)}]{Schroer2009}
\bibinfo{author}{\bibnamefont{{W. Schr\"oer and V. R. Vale}}},
  \bibinfo{journal}{J. Phys.: Condens. Matter} \textbf{\bibinfo{volume}{21}},
  \bibinfo{pages}{424119} (\bibinfo{year}{2009}).

\bibitem[{\citenamefont{{M. P. Allen and D. J. Tildesley}}(1987)}]{Allen}
\bibinfo{author}{\bibnamefont{{M. P. Allen and D. J. Tildesley}}},
  \emph{\bibinfo{title}{Computer Simulation of Liquids}}
  (\bibinfo{publisher}{Oxford University Press}, \bibinfo{year}{1987}).

\bibitem[{\citenamefont{{R. Fantoni, A. Giacometti, F. Sciortino, nd G.
  Pastore}}(2011)}]{Fantoni2011}
\bibinfo{author}{\bibnamefont{{R. Fantoni, A. Giacometti, F. Sciortino, nd G.
  Pastore}}}, \bibinfo{journal}{Soft Matter} \textbf{\bibinfo{volume}{7}},
  \bibinfo{pages}{2419} (\bibinfo{year}{2011}).

\bibitem[{\citenamefont{{R. Fantoni}}(2012{\natexlab{b}})}]{Fantoni2012}
\bibinfo{author}{\bibnamefont{{R. Fantoni}}}, \bibinfo{journal}{Eur. Phys. J.
  B} \textbf{\bibinfo{volume}{85}}, \bibinfo{pages}{108}
  (\bibinfo{year}{2012}{\natexlab{b}}).

\bibitem[{not()}]{note1}
\bibinfo{note}{Many different ways of defining a cluster have been proposed
  \cite{Lee1973,Ebeling1980,Gillan1983,Fisher1993,Friedman1979}, since the
  Bjerrum theory \cite{Bjerrum1926} of ionic associations first appeared. Our
  choice corresponds to the one of Gillan \cite{Gillan1983} and Caillol and
  Weis \cite{Caillol1995}.}

\bibitem[{\citenamefont{{L. Rovigatti, J. Russo, and F.
  Sciortino}}(2011)}]{Rovigatti2011}
\bibinfo{author}{\bibnamefont{{L. Rovigatti, J. Russo, and F. Sciortino}}},
  \bibinfo{journal}{Phys. Rev. Lett.} \textbf{\bibinfo{volume}{107}},
  \bibinfo{pages}{237801} (\bibinfo{year}{2011}).

\bibitem[{\citenamefont{{A. B. Bhatia and D. E. Thornton}}(1970)}]{Bhatia1970}
\bibinfo{author}{\bibnamefont{{A. B. Bhatia and D. E. Thornton}}},
  \bibinfo{journal}{Phys. Rev. B} \textbf{\bibinfo{volume}{2}},
  \bibinfo{pages}{3004} (\bibinfo{year}{1970}).

\bibitem[{\citenamefont{{R. Fantoni, D. Gazzillo, and A.
  Giacometti}}(2005)}]{Fantoni05b}
\bibinfo{author}{\bibnamefont{{R. Fantoni, D. Gazzillo, and A. Giacometti}}},
  \bibinfo{journal}{Phys. Rev. E} \textbf{\bibinfo{volume}{72}},
  \bibinfo{pages}{011503} (\bibinfo{year}{2005}).

\bibitem[{\citenamefont{{M. Rovere, D. W. Heermann, and K.
  Binder}}(1988)}]{Rovere1988}
\bibinfo{author}{\bibnamefont{{M. Rovere, D. W. Heermann, and K. Binder}}},
  \bibinfo{journal}{Europhys. Lett.} \textbf{\bibinfo{volume}{6}},
  \bibinfo{pages}{585} (\bibinfo{year}{1988}).

\bibitem[{\citenamefont{{M. Rovere, D. W. Heermann, and K.
  Binder}}(1990)}]{Rovere1990}
\bibinfo{author}{\bibnamefont{{M. Rovere, D. W. Heermann, and K. Binder}}},
  \bibinfo{journal}{J. Phys.: Condens. Matter} \textbf{\bibinfo{volume}{2}},
  \bibinfo{pages}{7009} (\bibinfo{year}{1990}).

\bibitem[{\citenamefont{{M. Rovere, P. Nielaba, and K.
  Binder}}(1993)}]{Rovere1993}
\bibinfo{author}{\bibnamefont{{M. Rovere, P. Nielaba, and K. Binder}}},
  \bibinfo{journal}{Z. Phys. B} \textbf{\bibinfo{volume}{90}},
  \bibinfo{pages}{215} (\bibinfo{year}{1993}).

\bibitem[{\citenamefont{{C. Vega, J. L. F. Abascal, C. McBride, and F.
  Bresme}}(2003)}]{Vega2003}
\bibinfo{author}{\bibnamefont{{C. Vega, J. L. F. Abascal, C. McBride, and F.
  Bresme}}}, \bibinfo{journal}{J. Chem. Phys.} \textbf{\bibinfo{volume}{119}},
  \bibinfo{pages}{964} (\bibinfo{year}{2003}).

\bibitem[{\citenamefont{{D. Frenkel and B. Smit}}(1996)}]{Frenkel-Smit}
\bibinfo{author}{\bibnamefont{{D. Frenkel and B. Smit}}},
  \emph{\bibinfo{title}{Understanding Molecular Simulation}}
  (\bibinfo{publisher}{Academic Press}, \bibinfo{address}{San Diego},
  \bibinfo{year}{1996}).

\bibitem[{\citenamefont{{A. Z. Panagiotopoulos}}(1987)}]{Panagiotopoulos87}
\bibinfo{author}{\bibnamefont{{A. Z. Panagiotopoulos}}}, \bibinfo{journal}{Mol.
  Phys.} \textbf{\bibinfo{volume}{61}}, \bibinfo{pages}{813}
  (\bibinfo{year}{1987}).

\bibitem[{\citenamefont{{A. Z. Panagiotopoulos, N. Quirke, M. Stapleton, and D.
  J. Tildesley}}(1988)}]{Panagiotopoulos88}
\bibinfo{author}{\bibnamefont{{A. Z. Panagiotopoulos, N. Quirke, M. Stapleton,
  and D. J. Tildesley}}}, \bibinfo{journal}{Mol. Phys.}
  \textbf{\bibinfo{volume}{63}}, \bibinfo{pages}{527} (\bibinfo{year}{1988}).

\bibitem[{\citenamefont{{B. Smit, Ph. De Smedt, and D.
  Frenkel}}(1989)}]{Smit89a}
\bibinfo{author}{\bibnamefont{{B. Smit, Ph. De Smedt, and D. Frenkel}}},
  \bibinfo{journal}{Mol. Phys.} \textbf{\bibinfo{volume}{68}},
  \bibinfo{pages}{931} (\bibinfo{year}{1989}).

\bibitem[{\citenamefont{{B. Smit and D. Frenkel}}(1989)}]{Smit89b}
\bibinfo{author}{\bibnamefont{{B. Smit and D. Frenkel}}},
  \bibinfo{journal}{Mol. Phys.} \textbf{\bibinfo{volume}{68}},
  \bibinfo{pages}{951} (\bibinfo{year}{1989}).

\bibitem[{\citenamefont{{J. V. Sengers and J. M. H.
  Levelt-Sengers}}(1978)}]{Sengers}
\bibinfo{author}{\bibnamefont{{J. V. Sengers and J. M. H. Levelt-Sengers}}}, in
  \emph{\bibinfo{booktitle}{Progress in Liquid Physics}}, edited by
  \bibinfo{editor}{\bibnamefont{{C. A. Croxton}}} (\bibinfo{publisher}{Wiley},
  \bibinfo{address}{Chichester}, \bibinfo{year}{1978}),
  chap.~\bibinfo{chapter}{4}.

\bibitem[{\citenamefont{{F. Wegner}}(1972)}]{Wegner1972}
\bibinfo{author}{\bibnamefont{{F. Wegner}}}, \bibinfo{journal}{Phys. Rev. B}
  \textbf{\bibinfo{volume}{5}}, \bibinfo{pages}{4529} (\bibinfo{year}{1972}).

\bibitem[{\citenamefont{{J.-M. Caillol, D. Levesque, and J. J.
  Weis}}(1996)}]{Caillol1996}
\bibinfo{author}{\bibnamefont{{J.-M. Caillol, D. Levesque, and J. J. Weis}}},
  \bibinfo{journal}{Phys. Rev. Lett.} \textbf{\bibinfo{volume}{77}},
  \bibinfo{pages}{4039} (\bibinfo{year}{1996}).

\bibitem[{\citenamefont{{J.-M. Caillol}}(2004)}]{Caillol2004}
\bibinfo{author}{\bibnamefont{{J.-M. Caillol}}}, \bibinfo{journal}{Condensed
  Matter Physics} \textbf{\bibinfo{volume}{7}}, \bibinfo{pages}{741}
  (\bibinfo{year}{2004}).

\bibitem[{\citenamefont{{J. V. Sengers and J. G. Shanks}}(2009)}]{Sengers2009}
\bibinfo{author}{\bibnamefont{{J. V. Sengers and J. G. Shanks}}},
  \bibinfo{journal}{J. Stat. Phys.} \textbf{\bibinfo{volume}{137}},
  \bibinfo{pages}{857} (\bibinfo{year}{2009}).

\bibitem[{\citenamefont{{C. Valeriani, P. J. Camp, J. W. Zwanikken, R. van
  Roij, and M. Dijkstra}}(2010)}]{Valeriani2010}
\bibinfo{author}{\bibnamefont{{C. Valeriani, P. J. Camp, J. W. Zwanikken, R.
  van Roij, and M. Dijkstra}}}, \bibinfo{journal}{Soft Matter}
  \textbf{\bibinfo{volume}{6}}, \bibinfo{pages}{2793} (\bibinfo{year}{2010}).

\bibitem[{\citenamefont{{A. Tani and D. Henderson}}(1983)}]{Tani1983}
\bibinfo{author}{\bibnamefont{{A. Tani and D. Henderson}}},
  \bibinfo{journal}{J. Chem. Phys.} \textbf{\bibinfo{volume}{79}},
  \bibinfo{pages}{2390} (\bibinfo{year}{1983}).

\bibitem[{\citenamefont{{J. K. Lee, J. A. Barker, and F. F.
  Abraham}}(1973)}]{Lee1973}
\bibinfo{author}{\bibnamefont{{J. K. Lee, J. A. Barker, and F. F. Abraham}}},
  \bibinfo{journal}{J. Chem. Phys.} \textbf{\bibinfo{volume}{58}},
  \bibinfo{pages}{3166} (\bibinfo{year}{1973}).

\bibitem[{\citenamefont{{W. Ebeling and M. Grigo}}(1980)}]{Ebeling1980}
\bibinfo{author}{\bibnamefont{{W. Ebeling and M. Grigo}}},
  \bibinfo{journal}{Am. Phys.} \textbf{\bibinfo{volume}{37}},
  \bibinfo{pages}{21} (\bibinfo{year}{1980}).

\bibitem[{\citenamefont{{H. L . Friedman and G. Larsen}}(1979)}]{Friedman1979}
\bibinfo{author}{\bibnamefont{{H. L . Friedman and G. Larsen}}},
  \bibinfo{journal}{J. Chem. Phys.} \textbf{\bibinfo{volume}{70}},
  \bibinfo{pages}{92} (\bibinfo{year}{1979}).

\bibitem[{\citenamefont{{N. Bjerrum}}(1926)}]{Bjerrum1926}
\bibinfo{author}{\bibnamefont{{N. Bjerrum}}}, \bibinfo{journal}{Kgl. Dan.
  Vidensk. Selsk. Mat.-fys. Medd.} \textbf{\bibinfo{volume}{7}},
  \bibinfo{pages}{1} (\bibinfo{year}{1926}).

\end{thebibliography}

\end{document}